\documentclass[11pt]{article}
\usepackage[totalwidth=460truept,totalheight=600truept]{geometry}
\usepackage{epsfig,latexsym}

\def\theequation{\arabic{section}.\arabic{equation}}

\renewcommand{\theequation}{\thesection.\arabic{equation}}
\global\arraycolsep=1truept
\input epsf
\input{tcilatex}

\begin{document}

\font\cmss=cmss10 \font\cmsss=cmss10 at 7pt \hfill \hfill IFUP-TH/04-08

\null 

\vspace{10pt}\vskip 1.5truecm

\begin{center}
\textbf{\Large DEFORMED DIMENSIONAL REGULARIZATION \\[0pt]
FOR ODD (AND\ EVEN)\ DIMENSIONAL\ THEORIES}

\bigskip \vskip 1truecm

\textsl{Damiano Anselmi}

\textit{Dipartimento di Fisica ``E. Fermi'', Universit\`{a} di Pisa, and INFN%
}
\end{center}

\vskip 2truecm

\begin{center}
\textbf{Abstract}
\end{center}

\bigskip

{\small I formulate a deformation of the dimensional-regularization
technique that is useful for theories where the common dimensional
regularization does not apply. The Dirac algebra is not dimensionally
continued, to avoid inconsistencies with the trace of an odd product of
gamma matrices in odd dimensions. The regularization is completed with an
evanescent higher-derivative deformation, which proves to be efficient in
practical computations. This technique is particularly convenient in three
dimensions for Chern-Simons gauge fields, two-component fermions and
four-fermion models in the large }$N${\small \ limit, eventually coupled
with quantum gravity. Differently from even dimensions, in odd dimensions it
is not always possible to have propagators with fully Lorentz invariant
denominators. The main features of the deformed technique are illustrated in
a set of sample calculations. The regularization is universal, local,
manifestly gauge-invariant and Lorentz invariant in the physical sector of
spacetime. In flat space power-like divergences are set to zero by default.
Infinitely many evanescent operators are automatically dropped.}

\vskip 1truecm

\vfill\eject

\section{Introduction}

\setcounter{equation}{0}

The dimensional-regularization technique \cite{bollini,thooft} is the most
efficient technique for the calculation of Feynman diagrams in quantum field
theory. When gauge bosons couple to fermions in a chiral invariant way,
gauge invariance is manifest. When gauge bosons couple to chiral currents,
the definition of $\gamma _{5}$ in even dimensions ($\gamma _{5}=\gamma
_{1}\gamma _{2}\gamma _{3}\gamma _{4}$ \cite{thooft,maison,collins}) breaks
the Lorentz symmetry in the dimensionally continued spacetime and generates
axial anomalies. Gauge invariance survives if and only if the one-loop gauge
anomalies vanish. This is a restriction on the matter content of the theory.
The Adler-Bardeen theorem \cite{adler} ensures that there exists a
subtraction scheme where anomalies vanish to all orders in perturbation
theory, once they vanish at one loop.

At the practical level, calculations with the dimensional-regularization
technique in parity violating theories are not more difficult than
calculations in parity invariant theories. The reason is that the presence
of $\gamma _{5}$ does not break the continued Lorentz symmetry in the
denominators of propagators, but only in the vertices and numerators of
propagators. Therefore, using appropriate projectors, the Feynman integrals
can be decomposed into a basis of fully Lorentz invariant integrals. The
complication introduced by $\gamma _{5}$ is only algebraic and can be easily
treated with calculators.

On the other hand, the ordinary dimensional regularization is not universal,
in the sense that there exist models that cannot be dimensionally
regularized in the ordinary framework. The dimensionally continued Dirac
algebra has the property that the trace of an odd product of gamma matrices
is always equal to zero (see for example \cite{collins}). However, if $D$
denotes the physical spacetime dimension and $d=D-\varepsilon $ is its
continuation, the trace of the product of $D$ gamma matrices should tend to
the epsilon tensor in the limit $\varepsilon \rightarrow 0$, when $D$ is
odd. The three-dimensional four-fermion model in the large $N$ limit is
another example of theory that cannot be regularized with an ordinary
dimensional continuation. This model is not power-counting renormalizable,
but becomes renormalizable in the large $N$ expansion (where $N$ is the
number of fermion copies), after the resummation of fermion bubbles \cite
{parisi}. Nevertheless, the effective propagator obtained after this
resummation originates $\Gamma [0]$s in subleading diagrams. Ways to
regulate these $\Gamma [0]$s have been already presented in ref.s \cite
{largeN,largeN2}. Here I consider a more general framework.

\bigskip

In this paper I show that although the dimensional-regularization technique
does not apply to every model in a naive way, there always exist
deformations of the dimensional-regularization technique that regularize a
theory consistently at each order of the loop expansion in a manifestly
gauge-invariant way (up to the known anomalies) and preserve Lorentz
invariance in the physical subsector of spacetime. These deformations are
obtained combining variants of the usual dimensional technique with
evanescent higher-derivative corrections, multiplied by an extra cut-off.

Both the dimensional and higher-derivative regularizations have virtues and
weak points. Fortunately, the weak points of the two techniques have an
empty intersection, so an appropriate combination of the two can enhance the
virtues of both.

The higher-derivative regularization is gauge invariant and in principle
universal, but it regulates only higher-loop divergences. One-loop
divergences have to be treated separately \cite{fadeevslavnov}. Using
appropriate Pauli-Villars\ fields, Fadeev and Slavnov have shown that it is
possible to regulate the one-loop divergences in a gauge-invariant way when
the theory contains non-Abelian gauge fields \cite{fadeevslavnov}.
Presumably, the construction can be extended to gravity. However,
calculations with the higher-derivative technique are cumbersome. In quantum
gravity and other non-renormalizable theories exponentials are necessary and
the large number of additional vertices makes computations hard. Moreover,
the higher-derivative technique produces power-like divergences (linear,
quadratic, etc.). In the presence of gravity the powers of the cut-off can
be arbitrarily high.

Power-like divergences are RG invariant, namely they do not depend on the
dynamical scale $\mu $, because only logarithms $\log \Lambda /\mu $ force
the introduction of the RG scale. Due to this, there always exists a
subtraction scheme where power-like divergences are absent. This scheme is
called ``classically conformal scheme'', because when the theory is
classically conformal (namely it does not contain masses, nor dimensionful
couplings at the classical level) no mass nor dimensionful coupling is
generated by renormalization. Then, the dynamically generated scale $\mu $
is the unique dimensionful constant of the theory at the quantum level. The
dimensional regularization automatically selects the classically conformal
scheme. In the other regularization frameworks it is possible to reach this
scheme manually fine-tuning the local counterterms.

In three dimensions the dimensional regularization is inconsistent if the
theory contains, for example, two-component fermions coupled with
Chern-Simons gauge fields. In this case it is incorrect to set the trace of
an odd product of gamma matrices to zero. Nevertheless, this inconsistency
does not show up at one-loop, because one-loop diagrams in odd dimensions
have no logarithmic divergence. Since power-like divergences can be ignored,
at least in the classically conformal subtraction scheme, this is equivalent
to say that the theory is convergent at one-loop.

Now, given that the difficulties of the higher-derivative regularization are
confined to one loop, while the difficulties of the dimensional technique
show up only beyond one-loop, it is reasonable to argue that a suitable
combination of the two techniques can provide a consistent and universal
regularization framework. A generic higher-derivative deformation, however,
is difficult to handle in practical computations. I show in a number of
examples that if the higher-derivative deformation is also evanescent,
calculations can still be done efficiently.

\bigskip

When the Standard Model is coupled with quantum gravity the dimensional
regularization breaks the continued \textit{local} Lorentz symmetry, because
of the presence of $\gamma _{5}$ in the interactions. A similar breaking
takes place in three-dimensional quantum gravity coupled with Chern-Simons
gauge fields, two-component fermions and so on. In ref. \cite{pap4SM} it is
shown that it possible to dimensionally regularize quantum gravity coupled
with parity violating matter in such a way that the propagators of the
graviton, its ghosts and auxiliary fields have fully Lorentz invariant
denominators. Instead of gauge-fixing the residual local Lorentz symmetry
choosing a symmetric vielbein, a derivative Lorentz gauge fixing of the form 
$\partial ^{\mu }\omega _{\mu }^{ab}$ and a clever use of the auxiliary
fields do the job. Combining the results of \cite{pap4SM} with the ones of
the present paper it is possible to extend the regularization studied here
to odd-dimensional quantum gravity coupled with parity-violating matter.

\bigskip

Recapitulating, the purpose of this paper is to study deformations of the
dimensional technique that regularize in a manifestly gauge-invariant way
also models where the usual formulation does not apply. I explore several
types of deformations and search for the one that makes calculations more
efficient. It turns out that the most convenient framework is the one in
which the higher-derivative deformation is also evanescent. Another
important point is that in odd dimensions, differently from even dimensions,
it is not always possible to have propagators with fully Lorentz invariant
denominators. This complicates the evaluation of integrals, but not too
much. I illustrate the evaluation of some standard diagrams to convince the
reader that computations are still reasonably doable. Another advantage of
the deformed technique, in even and odd dimensions, is that infinitely many
evanescent operators, that are present in the usual, undeformed approach,
are automatically dropped.

In the literature various alternative definitions of $\gamma _{5}$ and the $%
\varepsilon $ tensor have been proposed. I use the 't Hooft-Veltman
prescription, which is known to be fully consistent. To my knowledge, there
exists no definition of $\gamma _{5}$ that commutes with all $\gamma _{\mu }$%
s, is fully consistent and manifestly gauge invariant modulo the
Adler-Bell-Jackiw anomalies. It is out of the purposes of this paper to
review the history of alternative proposals.

It is worth to remind that evenescent operators do not affect the S-matrix,
but produce at most scheme changes. Some properties of the ``theory of
evanescent operators'' are collected in Collins' book \cite{collins}. More
recent references are \cite{harv,nier}.

\bigskip

There is a variety of reasons for which it is good to have manifestly
gauge-invariant regularization techniques. For example, the existence of a
regularization with these properties is useful to prove the absence of gauge
anomalies. In other approaches it is necessary to resort to lengthy
cohomological classifications \cite{qualcuno} or deal with explicit and
involved cut-off dependencies, as in the exact-renormalization-group
approach \cite{parmensi}. Other theoretical applications concern the study
of renormalizability and finiteness beyond power counting, for example the
construction of consistent irrelevant deformations of renormalizable
theories \cite{pap3} in even and odd dimensions, and three-dimensional
quantum gravity coupled with matter \cite{pap1}, which, under certain
conditions, can be quantized as a finite theory \cite{pap2}.

The plan of the paper is as follows. In section 2 I describe the technique
and its main features. I write the deformed actions that regularize Dirac
fermions, Chern-Simons gauge fields and gravity and study the structure of
the renormalized actions to all orders in perturbation theory. Then I
proceed with sample calculations: the vacuum polarization in section 3 and
the axial anomalies in section 4. In section 5 I study Chern-Simons gauge
theories coupled with two-component fermions in three dimensions, and work
out the vacuum polarization and the one-loop fermion self-energy. In section
6 I\ study the three-dimensional four-fermion models in the large N
expansion and analogous scalar models. In section 7 I give a general recipe
to perform the evanescent higher-derivative deformation of $SO(d)$ invariant
theories in flat space. In section 8 I prove that power-like divergences are
absent in flat space. In section 9 I\ collect the conclusions. In the
appendix I show how to calculate some useful integrals and comment on
alternative non-evanescent higher-derivative deformations.

I work in the Euclidean framework, so the integrals are already Wick
rotated. No information is lost, since divergences are the same in the
Euclidean and Minkowskian frameworks \cite{collins}. With an abuse of
language, I call the $SO(d)$, $SO(D)$ and $SO(-\varepsilon )$ Euclidean
invariances (continued, physical and evanescent, respectively) \textit{%
Lorentz} symmetries, since no confusion can arise.

\section{The technique}

If the continued gamma matrices satisfy the $d$-dimensional Dirac algebra $%
\{\gamma ^{a},\gamma ^{b}\}=2\delta ^{ab}$ a standard argument \cite{collins}
proves that the trace of an odd product of gamma matrices is necessarily
zero. It is useful to recall here the proof. Using the cyclicity of the
trace and the Dirac algebra we have immediately 
\begin{equation}
d~\mathrm{tr}[\gamma ^{a}]=\mathrm{tr}[\gamma ^{a}\gamma ^{e}\gamma _{e}]=%
\mathrm{tr}[\gamma _{e}\gamma ^{a}\gamma ^{e}]=2~\mathrm{tr}[\gamma ^{a}]-%
\mathrm{tr}[\gamma ^{a}\gamma _{e}\gamma ^{e}]=(2-d)~\mathrm{tr}[\gamma
^{a}],  \label{arg}
\end{equation}
whence it follows that $\mathrm{tr}[\gamma ^{a}]=0$. Next, consider 
\begin{equation}
d~\mathrm{tr}[\gamma ^{a}\gamma ^{b}\gamma ^{c}]=\mathrm{tr}[\gamma
^{a}\gamma ^{b}\gamma ^{c}\gamma ^{e}\gamma _{e}]=\mathrm{tr}[\gamma
_{e}\gamma ^{a}\gamma ^{b}\gamma ^{c}\gamma ^{e}].  \label{argu}
\end{equation}
Using $\mathrm{tr}[\gamma ^{a}]=0$ it follows that the tensor $\mathrm{tr}%
[\gamma ^{a}\gamma ^{b}\gamma ^{c}]$ is completely antisymmetric. After a
few manipulations (\ref{argu}) gives 
\begin{equation}
d~\mathrm{tr}[\gamma ^{a}\gamma ^{b}\gamma ^{c}]=(6-d)~\mathrm{tr}[\gamma
^{a}\gamma ^{b}\gamma ^{c}]\text{,}  \label{argu3}
\end{equation}
whence $\mathrm{tr}[\gamma ^{a}\gamma ^{b}\gamma ^{c}]=0$. Repeating the
argument, it is possible to prove that the trace of the product of an
arbitrary odd number of gamma matrices is equal to zero.

This fact is incompatible with the existence of $2^{[D/2]}$-component
spinors in odd $D$, because then the trace of the product of $D$ gamma
matrices must give the epsilon tensor in the physical limit $d\rightarrow D$%
. For example, in $D=3$ 
\[
\mathrm{tr}[\gamma ^{a}\gamma ^{b}\gamma ^{c}]\rightarrow 2i\varepsilon
^{abc}. 
\]
It is worth to observe that this odd-dimensional problem is essentially
different from the problem of $\gamma _{5}$ in even dimensions. The standard
definitions of $\gamma _{5}$ and the $\varepsilon $ tensor break the
continued Lorentz invariance \cite{thooft,collins,maison}, but do not affect
the $d$-dimensional Dirac algebra $\{\gamma ^{a},\gamma ^{b}\}=2\delta ^{ab}$%
. In odd dimensions, instead, the continued Lorentz symmetry should be
broken at the level of the Dirac algebra \footnote{%
I assume that the trace is cyclic. Non-cyclic trace functionals have been
studied in the literature \cite{kreimer}.}.

\subsection{Algebra of gamma matrices}

We have spacetime indices $\mu ,\nu ,\rho \ldots $ running from $1$ to $d$;
Lorentz indices $a,b,c\ldots $ running from $1$ to $d$; physical Lorentz
indices $\bar{a},\bar{b},\bar{c}\ldots $ running from $1$ to $D$; evanescent
Lorentz indices $\hat{a},\hat{b},\hat{c}\ldots $ running from $D$ to $d$
(with $D$ excluded). There exists an epsilon tensor $\varepsilon _{\bar{a}%
_{1}\cdots \bar{a}_{D}}$ in the physical portion of spacetime, defined as
usual, but there exists no epsilon tensor in the evanescent portion of
spacetime.

The algebra of gamma matrices is the tensor product of a physical ($D$
dimensional) Dirac algebra, and an evanescent ($-\varepsilon $ dimensional)
Dirac algebra. The two commute: 
\begin{equation}
\{\gamma ^{\bar{a}},\gamma ^{\bar{b}}\}=2\delta ^{\bar{a}\bar{b}},\qquad
\{\gamma ^{\hat{a}},\gamma ^{\hat{b}}\}=2\delta ^{\hat{a}\hat{b}},\qquad
[\gamma ^{\bar{a}},\gamma ^{\hat{b}}]=0,  \label{defordir}
\end{equation}
where $\bar{a}=1,\ldots D$ and $D<\hat{a}<d$ (here $\varepsilon $ can be
imagined to be real and negative).

Spinors have $2^{[D/2]-\varepsilon /2}=2^{[D/2]}\cdot 2^{-\varepsilon /2}$
components, where [$n$] denotes the integral part of $n$. Write $\psi ^{\bar{%
\alpha}\hat{\alpha}}$, where $\bar{\alpha}=1,\ldots 2^{[D/2]}$ is the
physical spinor index and $\hat{\alpha}=1,\ldots 2^{-\varepsilon /2}$ is the
evanescent spinor index. The gamma matrices $\gamma _{\bar{\alpha}\hat{\alpha%
},\bar{\beta}\hat{\beta}}^{\bar{a}}$ act as the usual $2^{[D/2]}\times
2^{[D/2]}$ Hermitean Dirac matrices $\overline{\gamma }_{\bar{\alpha}\bar{%
\beta}}^{\bar{a}}$ on the physical spinor indices and the identity on the
evanescent spinor indices: 
\[
\gamma _{\bar{\alpha}\hat{\alpha},\bar{\beta}\hat{\beta}}^{\bar{a}}=%
\overline{\gamma }_{\bar{\alpha}\bar{\beta}}^{\bar{a}}\delta _{\hat{\alpha}%
\hat{\beta}}. 
\]
The trace of an arbitrary product of matrices $\gamma ^{\bar{a}}$ follows
immediately from the definition. In particular, when $D=3$ the trace of the
product of three such matrices is $2^{1-\varepsilon /2}i$ times the epsilon
tensor. The matrix $\gamma _{5}$ is defined as $\gamma _{5}=\gamma
^{1}\cdots \gamma ^{D}$ (equal to 1 if $D$ is odd).

The gamma matrices $\gamma _{\bar{\alpha}\hat{\alpha},\bar{\beta}\hat{\beta}%
}^{\hat{a}}$ act as formal $2^{-\varepsilon /2}\times 2^{-\varepsilon /2}$
Hermitean Dirac matrices $\widehat{\gamma }_{\hat{\alpha}\hat{\beta}}^{\hat{a%
}}$ ($\{\widehat{\gamma }^{\hat{a}},\widehat{\gamma }^{\hat{b}}\}=2\delta ^{%
\hat{a}\hat{b}}$) on the evanescent spinor indices and the identity on the
physical spinor indices: 
\[
\gamma _{\bar{\alpha}\hat{\alpha},\bar{\beta}\hat{\beta}}^{\hat{a}}=\delta _{%
\bar{\alpha}\bar{\beta}}\widehat{\gamma }_{\hat{\alpha}\hat{\beta}}^{\hat{a}%
}. 
\]
The trace of an odd product of matrices $\gamma ^{\hat{a}}$ is zero because
of the arguments recalled above. The trace of an even product of these
matrices is defined in the usual way.

The so defined physical and evanescent gamma matrices clearly commute.
Moreover, the trace of a product of gamma matrices factorizes into the
product of a trace in the physical spinor indices and a trace in the
evanescent spinor indices.

The contradiction (\negthinspace \ref{argu3}) is avoided thanks of the
commutativity of the physical and evanescent Dirac matrices. Obviously, $%
\mathrm{tr}[\gamma ^{\bar{a}}\gamma ^{\bar{b}}\gamma ^{\bar{c}%
}]=2^{1-\varepsilon /2}i\varepsilon ^{\bar{a}\bar{b}\bar{c}}$ and 
\[
d~\mathrm{tr}[\gamma ^{\bar{a}}\gamma ^{\bar{b}}\gamma ^{\bar{c}}]=\mathrm{tr%
}[\gamma ^{\bar{a}}\gamma ^{\bar{b}}\gamma ^{\bar{c}}(\gamma ^{\bar{e}%
}\gamma _{\bar{e}}+\gamma ^{\hat{e}}\gamma _{\hat{e}})]=\mathrm{tr}[\gamma _{%
\bar{e}}\gamma ^{\bar{a}}\gamma ^{\bar{b}}\gamma ^{\bar{c}}\gamma ^{\bar{e}%
}]+\mathrm{tr}[\gamma _{\hat{e}}\gamma ^{\bar{a}}\gamma ^{\bar{b}}\gamma ^{%
\bar{c}}\gamma ^{\hat{e}}]. 
\]
The first piece is in $D=3$ and the usual manipulations prove that it is
equal to $(6-D)~\mathrm{tr}[\gamma ^{\bar{a}}\gamma ^{\bar{b}}\gamma ^{\bar{c%
}}]=3~\mathrm{tr}[\gamma ^{\bar{a}}\gamma ^{\bar{b}}\gamma ^{\bar{c}}]$. The
second piece, because of the \textit{commutation} rule $[\gamma ^{\bar{a}%
},\gamma ^{\hat{b}}]=0$, gives back $\mathrm{tr}[\gamma ^{\bar{a}}\gamma ^{%
\bar{b}}\gamma ^{\bar{c}}\gamma _{\hat{e}}\gamma ^{\hat{e}}]=-\varepsilon ~%
\mathrm{tr}[\gamma ^{\bar{a}}\gamma ^{\bar{b}}\gamma ^{\bar{c}}]$. In total, 
\[
d~\mathrm{tr}[\gamma ^{\bar{a}}\gamma ^{\bar{b}}\gamma ^{\bar{c}%
}]=(3-\varepsilon )~\mathrm{tr}[\gamma ^{\bar{a}}\gamma ^{\bar{b}}\gamma ^{%
\bar{c}}], 
\]
which is consistent. Finally, it is immediate to show that $\mathrm{tr}%
[\gamma ^{\bar{a}}\gamma ^{\bar{b}}\gamma ^{\hat{c}}]=\mathrm{tr}[\gamma ^{%
\bar{a}}\gamma ^{\hat{b}}\gamma ^{\hat{c}}]=\mathrm{tr}[\gamma ^{\hat{a}%
}\gamma ^{\hat{b}}\gamma ^{\hat{c}}]=0$.

In the deformed dimensional technique, the Dirac action can be regularized
without making use of the hatted Dirac matrices: see formula (\ref{a1})
below. Then the hatted Dirac matrices appear nowhere in Feynman rules and
diagrams. Fermion traces just get an extra factor $2^{-\varepsilon /2}$,
which does not change the physical results. In this case, it is consistent
to work directly with $2^{[D/2]}$-component spinors $\psi ^{\bar{\alpha}}$.
In practice, this amounts to identify $\gamma ^{\bar{a}}$ with $\overline{%
\gamma }_{\bar{\alpha}\bar{\beta}}^{\bar{a}}$, replace the matrices $\gamma
^{\hat{a}}$ with the identity and ignore the evanescent spinor indices $\hat{%
\alpha}\hat{\beta}...$

In this framework, which I adopt in the rest of the paper, the
renormalization structure of the theory simplifies considerably. For
example, infinitely many evanescent operators are automatically dropped. To
be more explicit, observe that the matrices $\gamma ^{\hat{a}}$ allow the
contruction of infinite sets of evanescent operators with the same
dimensionality. Examples are the four-fermion operators 
\begin{equation}
(\overline{\psi }\gamma _{\mu _{1}\cdots \mu _{n}}\psi )^{2},\qquad \qquad
n=1,2,\ldots ,  \label{acca}
\end{equation}
$\gamma _{\mu _{1}\cdots \mu _{n}}$ being the completely antisymmetric
product of $n$ gamma matrices. The operators (\ref{acca}) are evanescent for 
$n>D$, but not zero. Using the deformed dimensional regularization with no
evanescent spinor indices, the operators (\ref{acca}) are exactly zero when $%
n>D$. This deformed framework admits only a finite number of evanescent
operators with a given dimensionality. They are constructed with the
evanescent components $\widehat{k}$, $\widehat{A}$, $\widehat{g}$ of
momenta, gauge vectors and the metric tensor (if gravity is quantized).

An example of regularized Dirac action that does make use of the hatted
Dirac matrices is given in Appendix B, formula (\ref{a2}).

\subsection{Propagators and integrals}

Momenta are split into physical and evanescent components, $p_{a}=(p_{\bar{a}%
},p_{\hat{a}})$, where $a=(\bar{a},\hat{a})$. The regularized propagators
are chosen so that they tend to zero in the usual way when $\overline{p}^{2}$
tends to infinity and tend to zero in a power-like way also when $\widehat{p}%
^{2}$ tends to infinity. The typical behavior of a bosonic propagator
considered in this paper is 
\begin{equation}
\frac{1}{\overline{p}^{2}+(m+\widehat{p}^{2}/\Lambda )^{2}},  \label{tupy}
\end{equation}
where $\Lambda $ is the cut-off of the higher-derivative deformation. The
typical behavior of a fermionic propagator is, roughly speaking, the square
root of (\ref{tupy}). Alternative propagators are studied in the next
sections.

Integrals are split as 
\begin{equation}
\int \frac{\mathrm{d}^{d}p}{(2\pi )^{d}}f(p)=\int \frac{\mathrm{d}^{D}%
\overline{p}}{(2\pi )^{D}}\int \frac{\mathrm{d}^{-\varepsilon }\widehat{p}}{%
(2\pi )^{-\varepsilon }}f(\overline{p},\widehat{p})  \label{decompo}
\end{equation}
and can be calculated in the following way. First calculate the $%
-\varepsilon $ dimensional integral using the usual formulas of the
dimensional regularization. Then calculate the remaining $D$ dimensional
integral, using again the formulas of the dimensional regularization. The
final $D$ integral is well-defined (in the sense of the dimensional
regularization, with complex $\varepsilon $) even if $D$ is strictly integer.

Sometimes it is convenient to do the $D$ integral before the $-\varepsilon $
integral. This exchange can be rigorously done only after $D$ is temporarily
continued to complex values. The procedure is: analytically continue the
integral to complex $D$ (without touching the gamma matrices $\gamma ^{\bar{a%
}}$), exchange the $D$ integral with the $-\varepsilon $ integral, calculate
the $D$ integral, take the limit $D\rightarrow $integer, calculate the $%
-\varepsilon $ integral. (The last two steps can be freely interchanged.)
More details are given together with the examples.

The regularization is removed letting $\varepsilon $ tend to zero at fixed $%
\Lambda $ and then letting $\Lambda \rightarrow \infty $ (see the examples
for further details). To show that the regularization is a good
regularization, it is necessary to prove that a convergent integral gives
back the initial integral when the regularization is removed. The key
ingredient to prove this statement is the theorem (see for example \cite
{collins}) stating that if $\widehat{f}\left( \widehat{P}\right) $ is a
regular function of $\widehat{P}\equiv (\widehat{p}_{1},\ldots ,\widehat{p}%
_{L})$ tending to zero at least as $1/(\widehat{P}^{2})^{\gamma }$ for some $%
\gamma >0$ when $\widehat{P}^{2}\equiv \widehat{p}_{1}^{2}+\ldots +\widehat{p%
}_{L}^{2}\rightarrow \infty $, then 
\begin{equation}
\lim_{\varepsilon \rightarrow 0}\int \frac{\mathrm{d}^{-L\varepsilon }%
\widehat{P}}{(2\pi )^{-L\varepsilon }}~\widehat{f}\left( \widehat{P}\right) =%
\widehat{f}(0),\qquad \text{where }\frac{\mathrm{d}^{-L\varepsilon }\widehat{%
P}}{(2\pi )^{-L\varepsilon }}\equiv \prod_{i=1}^{L}\frac{\mathrm{d}%
^{-\varepsilon }\widehat{p}_{i}}{(2\pi )^{-\varepsilon }}.  \label{colla}
\end{equation}
In practice, when $\varepsilon $ tends to zero the $-\varepsilon $ integral
acts as a delta function projecting the function $\widehat{f}$ onto $%
\widehat{P}=0$.

Now, consider a completely convergent Feynman integral, namely the integral
associated with a Feynman diagram $G$ that is superficially convergent and
has no subdivergences, with $L$ loops, and assume that the propagators are
of the form (\ref{tupy}), or equivalent. Temporarily continue $D$ to complex
values, in the way explained above. The integral can be written as 
\begin{equation}
\int \frac{\mathrm{d}^{-L\varepsilon }\widehat{P}}{(2\pi )^{-L\varepsilon }}~%
\widehat{f}\left( \widehat{P}\right) ,\qquad \text{with }\widehat{f}\left( 
\widehat{P}\right) =\int \frac{\mathrm{d}^{LD}\overline{P}}{(2\pi )^{LD}}f(%
\overline{P},\widehat{P}),  \label{ucci}
\end{equation}
with obvious notation. The limit $\widehat{P}^{2}\rightarrow \infty $ of $%
\widehat{f}\left( \widehat{P}\right) $ is studied letting any subset $s_{p}$
of momenta $\widehat{p}_{1},\ldots ,\widehat{p}_{L}$ become large. Each $%
s_{p}$ is associated with a subgraph of $G$. Since, by assumption, every
subgraph has a negative degree of divergence, Weinberg's theorem \cite{weth}
ensures that there exists a $\gamma >0$ such that the $\widehat{f}\left( 
\widehat{P}\right) $ of (\ref{ucci}) tends to zero at least as $1/(\widehat{P%
}^{2})^{\gamma }$, when $\widehat{P}^{2}\rightarrow \infty $. This is true
also for the temporarily continued $D$ (i.e. for values of $D$ slightly
different from its physical, integer, value). Then formula (\ref{colla}) can
be used, so the limit $\varepsilon \rightarrow 0$ trivializes the $%
-\varepsilon $ integration and returns the initial convergent $\overline{P}$
integral 
\[
\int \frac{\mathrm{d}^{LD}\overline{P}}{(2\pi )^{LD}}f(\overline{P},0). 
\]

Now I prove that Feynman diagrams are indeed regularized. Consider a generic
diagram. The integrand is a polynomial $Q$ in momenta, times a certain
number $n$ of propagators (\ref{tupy}), that I denote with $P(k,m)$. Let $k$
denote loop momenta, while external momenta are not written explicitly.
First study the $\widehat{k}$-integrations. For $\widehat{k}$ large at fixed 
$\overline{k}$, 
\begin{equation}
f(\overline{k})\equiv \int \frac{\mathrm{d}^{-\varepsilon }\widehat{k}}{%
(2\pi )^{-\varepsilon }}[P(k,m)]^{n}Q(k)\sim \int \frac{\mathrm{d}%
^{-\varepsilon }\widehat{k}}{(2\pi )^{-\varepsilon }}\sum_{p}\frac{1}{%
\widehat{k}^{p}}\sim \sum_{p}\frac{\Gamma (p/2+\varepsilon /2)}{\Gamma (p/2)}%
,  \label{sepia}
\end{equation}
where $p$ are integers (coefficients multiplying the terms of the sum $%
\sum_{p}$ are understood). The factor $\Gamma (p/2)$ is common type of
factor in dimensional regularization and absolutely harmless, even when $%
p/2\leq 0$, because it appears in the denominator. The integral is
regularized because the gamma function appearing in the numerator has an
argument shifted by $\varepsilon /2$. Multiple integrals produce shifts by $%
q\varepsilon /2$, with $q$ positive integer.

A similar argument can be repeated for the behavior of the $\overline{k}$%
-integration, namely 
\[
\int \frac{\mathrm{d}^{D}\overline{k}}{(2\pi )^{D}}f(\overline{k}), 
\]
\textit{after} the $\widehat{k}$-integration. To study the large-$\overline{k%
}$ behavior of $f(\overline{k})$, rescale $\overline{k}$ by a factor $%
\lambda $ in $f(\overline{k})$. At the same time, rescale $\widehat{k}$ by a
factor $\sqrt{\lambda }$ in the $\widehat{k}$-integral that defines $f(%
\overline{k})$, see (\ref{sepia}). Then $f(\lambda \overline{k})\sim
\sum_{p}\lambda ^{-p/2-\varepsilon /2}$ and therefore 
\begin{equation}
\int \frac{\mathrm{d}^{D}\overline{k}}{(2\pi )^{D}}f(\overline{k})\sim
\sum_{p}\frac{\Gamma (p/4+\varepsilon /4-D/2)}{\Gamma (p/4+\varepsilon /4)}.
\label{sepia2}
\end{equation}
The gamma function in the numerator has an argument shifted by $\varepsilon
/4$, so the integral is regularized. Multiple integrals produce shifts by $%
q\varepsilon /4$, with $q$ positive integer.

\bigskip

The counterterms are local both in the physical and evanescent components of
the external momenta. To prove this, it is sufficient to observe that if the
propagators are of the form (\ref{tupy}) or equivalent, after a sufficient
number of differentiations with respect to the physical or evanescent
components of the external momenta, every integral gets a negative overall
degree of divergence. Therefore, once the subdivergences have been
inductively subtracted, (\ref{colla}) can be used and the result is finite.
This implies that the divergent part is polynomial both in the physical and
evanescent components of the external momenta.

\bigskip

Summarizing, the $\widehat{P}$ integral, combined with the $\varepsilon
\rightarrow 0$ limit, is just a sort of delta-function projecting onto $%
\widehat{P}=0$, eventually collecting poles in $\varepsilon $. In practice,
the evanescent sector of spacetime dresses Feynman diagrams with appropriate
(gauge-invariant) regularizing distributions. This emphasizes the
mathematical meaning of the regularization used here and its elegance.

\subsection{Dirac action}

The Dirac action 
\begin{equation}
\mathcal{L}_{\text{Dirac}}=\overline{\psi }\left( \overline{D\!\!\!\!\slash}%
+m\right) \psi  \label{notre}
\end{equation}
is trivialized by the regularization. Indeed, write 
\begin{equation}
\mathrm{Det}\left( \overline{D\!\!\!\!\slash}+m\right) =\exp \left( \int 
\mathrm{d}^{d}x\int \frac{\mathrm{d}^{d}p}{(2\pi )^{d}}\mathrm{tr\ln }\left[ 
\overline{\partial \!\!\!\slash}+m+i\overline{p\!\!\!\slash}+i\overline{%
A\!\!\!\slash}(x)\right] \right) .  \label{argo}
\end{equation}
The integrand does not depend on $\widehat{p}$ and the $\widehat{p}$%
-integral of 1 is zero in dimensional regularization. The point is that the
lagrangian (\ref{notre}) is incomplete in $d$ dimensions. It does not
provide a propagator behaving like the ``square root'' of (\ref{tupy}) or
equivalent.

The Dirac fields can be efficiently regularized with an extra non-chiral
evanescent higher-derivative term: 
\begin{equation}
\mathcal{L}_{\text{Dirac}}=\overline{\psi }\left( \overline{D\!\!\!\!\slash}%
+m-\frac{\widehat{D^{2}}}{\Lambda }\right) \psi ,\qquad \langle \psi (p)~%
\overline{\psi }(-p)\rangle _{\mathrm{free}}=\frac{-i\overline{p\!\!\!\slash}%
+m+\widehat{p}^{2}/\Lambda }{\overline{p}^{2}+(m+\widehat{p}^{2}/\Lambda
)^{2}}.  \label{a1}
\end{equation}
The renormalization is studied first taking $\varepsilon \rightarrow 0$ at
fixed $\Lambda $ and then letting $\Lambda \rightarrow \infty $. The
converse does not work, since the argument (\ref{argo}) shows that the naive
limit $\Lambda \rightarrow \infty $ at $\varepsilon \neq 0$ produces zero,
not the initial theory. For a better behavior of integrals, the sign of the
evanescent piece is related to the sign of $m$. Here $m$ is assumed to be
positive.

The regularized Dirac action (\ref{a1}) does not contain hatted Dirac
matrices, because the hatted kinetic term is higher-derivative. The hatted
Dirac matrices, which appear nowhere in Feynman rules and diagrams, can be
ignored and it is consistent to work with $2^{[D/2]}$-component spinors $%
\psi ^{\bar{\alpha}}$.

\subsection{Chern-Simons action}

Consider the Chern-Simons action, 
\begin{equation}
\mathcal{L}_{\text{ChS}}=-\frac{i}{2\alpha }\varepsilon _{\bar{a}\bar{b}\bar{%
c}}F^{\bar{a}\bar{b}}A^{\bar{c}}.  \label{chernsi}
\end{equation}
For the moment I\ restrict to Abelian gauge fields. Because of the epsilon
tensor, the evanescent components $A^{\hat{a}}$ of the gauge field do not
have a kinetic term. It is necessary to introduce a second cut-off $\Lambda $
and suitable higher-derivative terms. The simplest possibility, reported in
formula (\ref{luno}) of appendix B, is not very convenient for calculations.
It is more convenient to regularize the Chern-Simons gauge field with an
evanescent higher-derivative deformation, 
\begin{equation}
\mathcal{L}_{\text{ChS}}=-\frac{i}{2\alpha }\varepsilon _{\bar{\mu}\bar{\nu}%
\bar{\rho}}F^{\bar{\mu}\bar{\nu}}A^{\bar{\rho}}+\frac{1}{\alpha \Lambda }F_{%
\overline{\mu }\hat{\nu}}^{2}-\frac{1}{2\alpha \Lambda ^{3}}F_{\hat{\mu}\hat{%
\nu}}\widehat{\partial ^{2}}F_{\hat{\mu}\hat{\nu}}.  \label{lchs}
\end{equation}
The gauge-fixing term can be deformed accordingly: 
\begin{equation}
\mathcal{L}_{\text{gf}}=\lambda \left( \overline{\partial A}-\frac{1}{%
\Lambda ^{2}}\widehat{\partial ^{2}}\widehat{\partial A}\right) ^{2}.
\label{bora}
\end{equation}
The consequent ghost action reads 
\begin{equation}
\mathcal{L}_{\text{ghost}}=\overline{C}\left( -\overline{\partial ^{2}}+%
\frac{\widehat{\partial ^{2}}^{2}}{\Lambda ^{2}}\right) C\text{.}
\label{borabora}
\end{equation}
The propagators are 
\begin{eqnarray}
\!\!\!\!\!\!\!\!\!\!\!\!\langle A_{\mu }(p)~A_{\nu }(-p)\rangle _{\mathrm{%
free}} &=&\frac{\alpha }{2(\overline{p}^{2}+\widehat{p}^{4}/\Lambda ^{2})}%
\left[ \varepsilon _{\bar{\mu}\bar{\nu}\bar{\rho}}p_{\bar{\rho}}+\frac{%
\widehat{p}^{2}}{\Lambda }\delta _{\bar{\mu}\bar{\nu}}+\Lambda \delta _{\hat{%
\mu}\hat{\nu}}+\frac{p_{\mu }p_{\nu }\left( 1/(\alpha \lambda )-\widehat{p}%
^{2}/\Lambda \right) }{\overline{p}^{2}+\widehat{p}^{4}/\Lambda ^{2}}\right]
,  \nonumber \\
\langle C(p)~\overline{C}(-p)\rangle _{\mathrm{free}} &=&\frac{1}{\overline{p%
}^{2}+\widehat{p}^{4}/\Lambda ^{2}}.  \label{LG}
\end{eqnarray}
All denominators have the same structure as the Dirac propagator (\ref{a1})
at $m=0$. It is not possible to have $SO(d)$ invariant denominators.
Nevertheless, since denominators have a coherent structure (up to masses)
the use Feynman parameters in calculations is efficient.

The generalization to non-Abelian gauge fields is straightforward, 
\[
\mathcal{L}_{\text{ChS}}=-\frac{i}{2\alpha }\varepsilon _{\bar{\mu}\bar{\nu}%
\bar{\rho}}\left( F_{\bar{\mu}\bar{\nu}}^{i}A_{\bar{\rho}}^{i}-\frac{1}{3}%
f_{ijk}A_{\bar{\mu}}^{i}A_{\bar{\nu}}^{j}A_{\bar{\rho}}^{k}\right) +\frac{1}{%
\alpha \Lambda }(F_{\bar{\mu}\hat{\nu}}^{i})^{2}-\frac{1}{2\alpha \Lambda
^{3}}F_{\hat{\mu}\hat{\nu}}^{i}\widehat{D^{2}}^{ij}F_{\hat{\mu}\hat{\nu}%
}^{j}, 
\]
where $i,j,k$ are indices of the adjoint representation of the gauge group.
The propagators are of course the same as before.

\subsection{Gravity}

For completeness, I comment on the regularization of gravity coupled with
parity-violating matter. Here an additional problem appears: the Lorentz
symmetry is a local symmetry, so the breaking of $SO(d)$ to $SO(D)\otimes
SO(-\varepsilon )$ is more delicate. For example, it is not possible to
gauge-fix the local Lorentz symmetry using the symmetric gauge. The
quadratic part of the Einstein lagrangian 
\begin{equation}
\mathcal{L}=\frac{1}{2\kappa ^{2}}\sqrt{g}R  \label{ll}
\end{equation}
does not depend on the antisymmetric part of the quantum fluctuation $%
\widetilde{\phi }_{\mu a}=e_{\mu a}-\delta _{\mu a}$, for which additional
terms must be provided. An economical arrangement is worked out in ref. \cite
{pap4SM}. Start from the ordinary, $SO(d)$ invariant situation and gauge-fix
diffeomorphisms and the Lorentz symmetry with the gauge-fixings $\partial
_{\mu }(\sqrt{g}g^{\mu \nu })$ and $\mathcal{D}^{\mu }\omega _{\mu }{}^{ab}$%
, respectively. The gauge-fixing terms are 
\begin{equation}
\mathcal{L}_{\mathrm{gf}}=\frac{1}{2\lambda }(\partial _{\mu }\sqrt{g}g^{\mu
\nu })^{2}+\frac{1}{2\xi }\sqrt{g}(\mathcal{D}^{\mu }\omega _{\mu
}{}^{ab})^{2}.  \label{ghefa}
\end{equation}
The first term is Lorentz invariant, the second is diffeomorphism invariant.
This diagonalizes the ghost action. In particular, the propagator of the
Lorentz ghosts $C^{ab}$ is just $1/p^{2}$ times the identity.

Now, decompose the second term of (\ref{ghefa}) as 
\begin{equation}
\frac{1}{2\xi }\sqrt{g}\left[ (\widetilde{\mathcal{D}}^{\mu }\omega _{\mu
}{}^{\bar{a}\bar{b}})^{2}+(\widetilde{\mathcal{D}}^{\mu }\omega _{\mu }{}^{%
\hat{a}\hat{b}})^{2}+2(\widetilde{\mathcal{D}}^{\mu }\omega _{\mu }{}^{\bar{a%
}\hat{b}})^{2}\right]  \label{interpreta}
\end{equation}
and interpret this decomposition in the context of the theory with broken
Lorentz symmetry. The covariant derivatives $\mathcal{D}_{\mu }$ have been
replaced by new covariant derivatives $\widetilde{\mathcal{D}}_{\mu }$,
defined with the spin connections $\omega _{\mu }{}^{\bar{a}\bar{b}},$ $%
\omega _{\mu }{}^{\hat{a}\hat{b}}$ of the reduced Lorentz group $%
SO(D)\otimes SO(-\varepsilon )$. The first two terms of (\ref{interpreta})
can be viewed as the gauge-fixings of $SO(D)\otimes SO(-\varepsilon )$. The
third term of (\ref{interpreta}) can be viewed as an addition to the
lagrangian (\ref{ll}), that provides the missing propagator for the
antisymmetric part of $\widetilde{\phi }_{\mu a}$. This addition is
legitimate, because it is a scalar density under diffeomorphisms, a scalar
under $SO(D)\otimes SO(-\varepsilon )$ rotations, and a true regularization
term, in the sense that it formally disappears in the physical limit $%
d\rightarrow D$. The rearrangement is 
\begin{equation}
\mathcal{L}^{\prime }=\frac{1}{2\kappa ^{2}}\sqrt{g}R+\frac{1}{\xi }\sqrt{g}(%
\widetilde{\mathcal{D}}^{\mu }\omega _{\mu }{}^{\bar{a}\hat{b}})^{2},\qquad 
\mathcal{L}_{\text{gf}}^{\prime }=\frac{1}{2\lambda }(\partial _{\mu }\sqrt{g%
}g^{\mu \nu })^{2}+\frac{1}{2\xi }\sqrt{g}\left[ (\widetilde{\mathcal{D}}%
^{\mu }\omega _{\mu }{}^{\bar{a}\bar{b}})^{2}+(\widetilde{\mathcal{D}}^{\mu
}\omega _{\mu }{}^{\hat{a}\hat{b}})^{2}\right] .  \label{high}
\end{equation}
The quadratic part of the sum $\mathcal{L}^{\prime }+\mathcal{L}_{\text{gf}%
}^{\prime }$ is clearly equal to the quadratic part of $\mathcal{L}+\mathcal{%
L}_{\text{gf}}$ . This ensures that the propagator of $\widetilde{\phi }%
_{\mu a}$ is unmodified and in particular its denominators are $SO(d)$
invariant. The difference $\mathcal{L}^{\prime }+\mathcal{L}_{\text{gf}%
}^{\prime }-\mathcal{L}-\mathcal{L}_{\text{gf}}$ is made of cubic terms, due
to the difference between the covariant derivatives $\mathcal{D}_{\mu }$ and 
$\widetilde{\mathcal{D}}_{\mu }$. Another modification is in the ghost
action, where the mixed Lorentz ghosts $C^{\bar{a}\hat{b}}$ are suppressed,
as well as the companion antighosts. The propagator of the reduced Lorentz
ghosts is still $1/p^{2}$ times the identity. Complete details (and a way to
avoid certain IR nuisances due to the higher-derivative gauge fixing) can be
found in ref. \cite{pap4SM}.

In summary, there exists a way to dimensionally regularize gravity coupled
with the Standard Model or, in general, parity violating matter, in such a
way that all propagators have $SO(d)$ invariant denominators. This property
simplifies perturbative calculations.

\subsection{Renormalization structure and stability of the deformed actions}

Here I study the structure of the renormalized action in the deformed
dimensional-regularization framework at arbitrarily high orders in
perturbation theory.

According to renormalization theory, every allowed counterterm should appear
in the renormalized action, multiplied by an independent coupling. The
coupling has to run appropriately, to ensure RG invariance. Evanescent
operators are an exception to this rule. This is important for the following
reason.

The deformed actions (\ref{a1}), (\ref{lchs}), (\ref{high}) and the ones
discussed in the next sections produce convenient propagators for efficient
perturbative calculations. However, those actions are written with \textit{%
ad hoc} deformations. For example, the coefficients of the two Chern-Simons
deformations in (\ref{lchs}) are related to each other in such a way to
produce the nice propagator (\ref{LG}). If the deformations in (\ref{lchs})
were multiplied by independent parameters, then the propagator would be much
more complicated.

Now, the evanescent sector of the theory does not mix into the physical
sector \cite{collins,harv}. This means that evanescent operators do not
affect the S matrix and the physical correlation functions, but produce at
most scheme changes. So, it is unnecessary to multiply the evanescent
operators by new independent couplings: evanescent counterterms, such as 
\[
\frac{1}{\varepsilon }\overline{\psi }\widehat{D\!\!\!\!\slash}\psi ,\qquad 
\frac{1}{\varepsilon }\frac{1}{\Lambda }F_{\overline{\mu }\hat{\nu}%
}^{2},\qquad \frac{1}{\varepsilon }\frac{1}{\Lambda }F_{\hat{\mu}\hat{\nu}%
}^{2},\qquad \frac{1}{\varepsilon }\sqrt{g}(\omega _{\mu }{}^{\bar{a}\hat{b}%
})^{2}, 
\]
etc., can be subtracted just as they come, at higher orders (starting from
one loop). This procedure violates RG invariance in physical correlation
functions only by contributions that vanish in the physical limit $%
\varepsilon \rightarrow 0$ and therefore have no physical significance.

Concluding, the complete renormalized action $\mathcal{L}_{\mathrm{R}}$ has
a non-evanescent sector $\mathcal{L}_{\text{non-ev}}$ and an evanescent
sector $\mathcal{L}_{\text{ev}}$. The non-evanescent sector has the usual
structure: it contains renormalization constants for every field and
coupling, and every allowed non-evanescent term is multiplied by independent
parameters. The structure of the evanescent sector, instead, is completely
free. Simbolically, 
\begin{equation}
\mathcal{L}_{\mathrm{R}}=\mathcal{L}_{\text{non-ev}}[Z_{\phi }^{1/2}\phi
,\lambda Z_{\lambda }\mu ^{p\varepsilon }]+\mathcal{L}_{\text{ev}}[\phi
,\varepsilon ].  \label{tora}
\end{equation}
In particular, the structures of the evanescent sectors in (\ref{a1}), (\ref
{lchs}) and (\ref{high}) are \textit{tree-level} structures and do not need
to be preserved at higher orders. It is possible to carry out every
calculation with the propagators produced by the tree-level structures (\ref
{a1}), (\ref{lchs}) and (\ref{high}) and subtract the evanescent
counterterms, at higher orders, just as they come.

\section{Sample calculation: vacuum polarization}

\setcounter{equation}{0}

In this and the next sections I illustrate the calculation of Feynman
diagrams with the deformed technique. I start from fermions coupled with
external gauge fields in four dimensions.

Consider the lagrangian (\ref{a1}). There are two vertices, with one or two
photon legs, 
\begin{equation}
\begin{tabular}{cc}
\hbox{\epsfig{figure=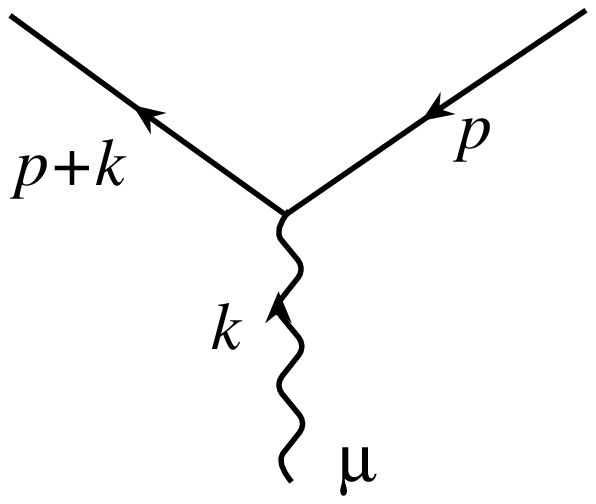,height=3cm,width=4cm}} & %
\hbox{\epsfig{figure=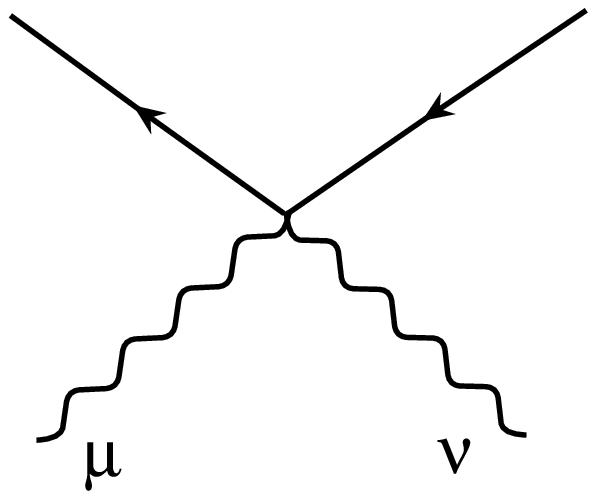,height=3cm,width=4cm}} \\ 
$-ie\mu ^{\varepsilon /2}\overline{\gamma }_{\mu }+{\frac{e\mu ^{\varepsilon
/2}}{\Lambda }}(2\widehat{p}+\widehat{k})_{\mu }$ & $-2e^{2}\mu
^{\varepsilon }\widehat{\delta }_{\mu \nu }/\Lambda $%
\end{tabular}
\label{vertices}
\end{equation}
The one-loop vacuum polarization reads 
\begin{equation}
VP=e^{2}\mu ^{\varepsilon }\int \frac{\mathrm{d}^{D}\overline{p}}{(2\pi )^{D}%
}\frac{2^{\varepsilon }\pi ^{\varepsilon /2}}{\Gamma (-\varepsilon /2)}%
\int_{0}^{\infty }t^{-1-\varepsilon /2}~\mathrm{d}t\frac{\mathrm{tr}[\gamma
_{\bar{a}}(-i\overline{p\!\!\!\slash}+\widetilde{t}/\Lambda )\gamma _{\bar{b}%
}(-i\overline{p\!\!\!\slash}+i\overline{k\!\!\!\slash}+\widetilde{t}/\Lambda
)]}{(\overline{p}^{2}+\widetilde{t}^{2}/\Lambda ^{2})\left( (\overline{p}%
-k)^{2}+\widetilde{t}^{2}/\Lambda ^{2}\right) },  \label{self}
\end{equation}
where $\widetilde{t}=t+m\Lambda $ and $t=\widehat{p}^{2}$. The external
indices and momenta have been projected onto the physical spacetime (e.g. $%
\widehat{k}=0$), so the evanescent contributions to the vertices can be
ignored in the calculation.

\bigskip

In the intermediate steps of a calculation, it is convenient to continue $D$
to complex values, and later replace it with its integer value. The general
recipe it as follows. First work out the numerators using the Dirac-algebra
conventions of section 2, and keep $D$ equal to its physical, therefore
integer, value. The integrand is a tensor depending on internal and external
momenta. Using several projectors, constructed with $\delta _{\bar{b}}^{\bar{%
a}}$, $\delta _{\hat{b}}^{\hat{a}}$, the epsilon tensor and the external
momenta, decompose the integral into the sum of a certain number of scalar
integrals, multiplied by suitable projectors. The scalar integrals can be
calculated separately. In these integrals, continue $D$ to complex values,
using the conventions of the dimensional regularization (the continuation is
straightforward, at this point). Calculate the $D$-integral and the $t$
integral (the order of these integrations is not crucial after the $D$
continuation) and then let $D$ tend back to its physical value. The
properties of the deformed regularization ensure that after the $t$ integral
this limit is smooth. The intermediate continuation to complex $D$s does not
touch the physical Dirac algebra and so avoids the inconsistencies mentioned
in section 2.

In the example (\ref{self}), working out the Dirac algebra in $D=4$ we
obtain $VP=Ak_{\bar{a}}k_{\bar{b}}+Bk^{2}\delta _{\bar{a}\bar{b}}$, where $A$
and $B$ are certain scalar integrals. It is immediate to prove gauge
invariance, namely $A+B=0$, and it remains to calculate 
\begin{equation}
(DB+A)k^{2}=\frac{2^{[D/2]}e^{2}~2^{\varepsilon }\pi ^{\varepsilon /2}\mu
^{\varepsilon }}{\Gamma (-\varepsilon /2)}\int \frac{\mathrm{d}^{D}\overline{%
p}}{(2\pi )^{D}}\int_{0}^{\infty }t^{-1-\varepsilon /2}~\mathrm{d}t\frac{%
(D-2)\overline{p}\cdot (\overline{p}-k)+D\widetilde{t}^{2}/\Lambda ^{2}}{(%
\overline{p}^{2}+\widetilde{t}^{2}/\Lambda ^{2})\left( (\overline{p}-k)^{2}+%
\widetilde{t}^{2}/\Lambda ^{2}\right) }.  \label{scaff}
\end{equation}

Now I illustrate some ways to calculate (\ref{scaff}). It is tempting to do
the $t$ integral immediately, using the formula (\ref{mil}) of the appendix,
but this procedure is not efficient. The result is for $D=4$ and $m=0,$%
\begin{equation}
3Bk^{2}=\frac{2^{1+\varepsilon }\pi ^{1+\varepsilon /2}e^{2}\mu
^{\varepsilon }\Lambda ^{-\varepsilon /2}}{\Gamma (-\varepsilon /2)\sin (\pi
\varepsilon /4)}\int \frac{\mathrm{d}^{4}\overline{p}}{(2\pi )^{4}}\frac{%
(3a-b+k^{2})ba^{-\varepsilon /4}-(3b-a+k^{2})ab^{-\varepsilon /4}}{ab(b-a)},
\label{bona}
\end{equation}
where $a=\overline{p}^{2}$ and $b=(\overline{p}-k)^{2}$. The $\overline{p}$
integration in (\ref{bona}) is hard to do, although it is always possible to
work out the divergent part of this expression taking two derivatives with
respect to the external momentum $k$ and then setting $k$ to zero.

Instead of doing the $t$ integral immediately, it is better to temporarily
continue to complex $D$ as explained above, use Feynman parameters, and then
integrate over $\overline{p}$ and $t$ in any preferred order. The $D$
integration in (\ref{scaff}) gives 
\begin{equation}
(D-1)Bk^{2}=-\frac{e^{2}~\mu ^{\varepsilon }\Lambda ^{-\varepsilon
/2}(D-1)\Gamma (2-D/2)}{2^{D-[D/2]-1-\varepsilon }\pi ^{D/2-\varepsilon
/2}\Gamma (-\varepsilon /2)}\int_{0}^{1}\mathrm{d}x~k_{x}^{2}\int_{0}^{%
\infty }t^{-1-\varepsilon /2}~\mathrm{d}t~\left[ k_{x}^{2}+(t+m)^{2}\right]
^{D/2-2},  \label{genericD2}
\end{equation}
after a rescaling of $t$. It is convenient to take the $D\rightarrow 4$
limit before integrating over $t$ and $x$. The spurious divergence
proportional to $\Gamma (2-D/2)$ is killed by the $t$ integral, as promised.
There does not exist a domain of (complex) values for $\varepsilon $ where
the $t$ integral is convergent, as it stands. It is necessary to split it
into a finite sum of integrals that separately admit convergence domains.
This is done in the appendix. The expansion in powers of $\varepsilon $ can
be studied with the help of formula (\ref{cicio}): 
\[
B\frac{\pi ^{2}}{e^{2}}=-\frac{1}{3}\left( \frac{1}{\varepsilon }-\frac{1}{2}%
\ln \frac{\Lambda }{4\pi \mu }\right) +\frac{1}{6}\gamma _{E}+\frac{1}{2}%
\int_{0}^{1}\mathrm{d}x~x(1-x)\ln \left[ x(1-x)k^{2}/\mu ^{2}+m^{2}/\mu ^{2}%
\right] . 
\]
After the identification $1/\varepsilon \sim \ln \Lambda /\mu $, this
expression agrees with the known one \cite{peskin}, up to a change of scheme.

\section{Axial anomalies}

\setcounter{equation}{0}

Now I\ illustrate the calculation of anomalies with the deformed technique,
using the regularized lagrangian (\ref{a1}). The axial transformation $%
\delta _{5}\psi =i\alpha \gamma _{5}\psi $, $\delta _{5}\overline{\psi }%
=i\alpha \overline{\psi }\gamma _{5}$ is associated with the current $J_{5}^{%
\bar{a}}=\overline{\psi }\gamma _{5}\gamma ^{\bar{a}}\psi $. Using the field
equations of (\ref{a1}) at $m=0$, which are $\overline{D\!\!\!\!\slash}\psi =%
\widehat{D^{2}}\psi /\Lambda $, the divergence of the axial current equals $%
\partial _{\bar{a}}J_{5}^{\bar{a}}=2\overline{\psi }\gamma _{5}\widehat{D^{2}%
}\psi /\Lambda $. The axial anomaly is 
\[
\mathcal{A}=\left\langle \partial _{\bar{a}}J_{5}^{\bar{a}}\right\rangle =%
\frac{2}{\Lambda }\left\langle \overline{\psi }\gamma _{5}\widehat{D^{2}}%
\psi \right\rangle =-\frac{2}{\Lambda }\mathrm{Tr}\!\!\left[ \gamma _{5}%
\widehat{D^{2}}\frac{1}{\overline{D\!\!\!\!\slash}-\widehat{D^{2}}/\Lambda }%
\right] . 
\]
The evanescent part $A_{\hat{\mu}}$ of the gauge vector can be set to zero,
since here it appears only as an external leg. This amounts to replace $%
\widehat{D^{2}}$ with minus the squared evanescent momentum $\widehat{p}%
^{2}=t$. To calculate the trace Tr it is convenient to choose a basis of
plane waves and use e$^{-ipx}\partial _{\mu }$e$^{ipx}=\partial _{\mu
}+ip_{\mu }$, obtaining 
\begin{equation}
\mathcal{A}=\frac{2^{\varepsilon }\pi ^{\varepsilon /2}}{\Lambda \Gamma
(-\varepsilon /2)}\int \frac{\mathrm{d}^{D}\overline{p}}{(2\pi )^{D}}%
\int_{0}^{\infty }t^{-\varepsilon /2}~\mathrm{d}t~\sum_{n=0}^{\infty }%
\mathrm{tr}\!\!\left[ \gamma _{5}\frac{1}{i\overline{p\!\!\!\slash}+%
\overline{\partial \!\!\!\slash}+t/\Lambda }\left( \frac{-ie\mu
^{\varepsilon /2}}{i\overline{p\!\!\!\slash}+\overline{\partial \!\!\!\slash}%
+t/\Lambda }A\!\!\!\slash\right) ^{n}\right] .  \label{enne}
\end{equation}
The denominator has been expanded in powers of the gauge field. It is
understood that the power $(AB)^{n}$, with $A$ and $B$ non-commuting
operators, is a symbolic notation to denote the product $ABABAB$... ($n$
times). Only the terms with $n=2,3,4$ can give non-vanishing contributions.
In the limit $\varepsilon \rightarrow 0$ the contributions with $n>4$ are
killed by the factor $1/\Gamma (-\varepsilon /2)$, since the $t$-$\overline{p%
}$ integral is convergent in that case. If the gauge fields are Abelian,
only $n=2$ gives a non-vanishing contribution.

The trace of (\ref{enne}) is calculated strictly in four dimensions,
according to the prescriptions of the deformed regularization technique. For 
$n=2$ we have immediately 
\[
\mathcal{A}=\frac{2^{2+\varepsilon }e^{2}\pi ^{\varepsilon /2}\mu
^{\varepsilon }\Lambda ^{-\varepsilon /2}}{\Gamma (-\varepsilon /2)}%
\varepsilon _{\bar{a}\bar{b}\bar{c}\bar{d}}k_{1}^{\bar{a}}A_{1}^{\bar{b}}k^{%
\bar{c}}A_{2}^{\bar{d}}\int \frac{\mathrm{d}^{D}\overline{p}}{(2\pi )^{D}}%
\int_{0}^{\infty }\frac{s^{-\varepsilon /4}~\mathrm{d}s}{((\overline{p}%
+k_{1})^{2}+s)(\overline{p}^{2}+s)((\overline{p}-k_{2})^{2}+s)}. 
\]
Using Feynman parameters, it is convenient to integrate first over $%
\overline{p}$, then over $s$. The $\overline{p}$ integral is already
convergent, so here there is no need to keep $D$ different from 4 in the
intermediate steps. The $s$ can be done using (\ref{mil}). After these two
integrations the limit $\varepsilon \rightarrow 0$ gives immediately the
known result, 
\[
\mathcal{A}=\left\langle \partial _{\bar{a}}J_{5}^{\bar{a}}\right\rangle =-%
\frac{e^{2}}{4\pi ^{2}}\varepsilon _{\bar{a}\bar{b}\bar{c}\bar{d}}k_{1}^{%
\bar{a}}A_{1}^{\bar{b}}k^{\bar{c}}A_{2}^{\bar{d}}=\frac{e^{2}}{16\pi ^{2}}%
\varepsilon _{\bar{a}\bar{b}\bar{c}\bar{d}}F^{\bar{a}\bar{b}}F^{\bar{c}\bar{d%
}}. 
\]

Summarizing, the calculation of anomalies with the deformed technique is not
more complicated than with the usual technique. It is simpler at the level
of algebraic manipulations of numerators, because the Dirac algebra stays in
the physical spacetime. On the other hand, the deformed calculation involves
a splitting of integrations. The regularization is due to the integration
over a sort of (squared) mass $s$.

\section{Chern-Simons theories in flat space}

\setcounter{equation}{0}

I consider Abelian Chern-Simons $U(1)$ gauge theory coupled with
two-component fermions in three dimensions, 
\begin{equation}
\mathcal{L}=-\frac{i}{2\alpha }\varepsilon _{\mu \nu \rho }F^{\mu \nu
}A^{\rho }+\overline{\psi }(\overline{D\!\!\!\!\slash}+m)\psi .  \label{laCS}
\end{equation}
This example is instructive, because the known dimensional-regularization
techniques do not apply. The regularized gauge-fixed lagrangian is the sum
of (\ref{lchs}) plus (\ref{bora}) plus (\ref{borabora}) plus (\ref{a1}).

\bigskip

\textbf{Vacuum polarization.} The vacuum polarization is made of two
contributions: one can be derived directly from formula (\ref{genericD2})
setting $D=3$; the second contribution comes from the trace of the product
of three gamma matrices in (\ref{self}).

The first contribution is of the form $Ak_{\bar{a}}k_{\bar{b}}+Bk^{2}\delta
_{\bar{a}\bar{b}}$. Multiplying by $k_{\bar{b}}$ it is immediate to prove
that $A+B=0$. The trace gives, from (\ref{genericD2})

\begin{equation}
2Bk^{2}=-\frac{e^{2}~\mu ^{\varepsilon }\Lambda ^{-\varepsilon /2}}{%
2^{-\varepsilon }\pi ^{1-\varepsilon /2}\Gamma (-\varepsilon /2)}\int_{0}^{1}%
\mathrm{d}x~k_{x}^{2}\int_{0}^{\infty }t^{-1-\varepsilon /2}~\mathrm{d}t~%
\left[ k_{x}^{2}+(t+m)^{2}\right] ^{-1/2}.  \label{alla}
\end{equation}
The $t$ integral can be evaluated with the technique explained in the
appendix, formula (\ref{appa}), and gives a certain hypergeometric function.
After taking the $\varepsilon \rightarrow 0$ limit, the integration over $x$
is immediate.

The second contribution is, after the $\overline{p}$ integration, 
\[
-\frac{2^{\varepsilon }\pi ^{\varepsilon /2}e^{2}~\mu ^{\varepsilon }\Lambda
^{-\varepsilon /2}}{4\pi \Gamma (-\varepsilon /2)}\varepsilon _{\bar{a}\bar{b%
}\bar{c}}k^{\bar{c}}\int_{0}^{1}\mathrm{d}x\int_{0}^{\infty
}t^{-1-\varepsilon /2}(t+m)~\mathrm{d}t~\left[ k_{x}^{2}+(t+m)^{2}\right]
^{-1/2} 
\]
and can be worked out in the same way as (\ref{alla}).

The final result is \cite{rosen,richter} 
\begin{equation}
\left. VP\right| _{D=3}=-\frac{e^{2}(k^{2}\delta _{\bar{a}\bar{b}}-k_{\bar{a}%
}k_{\bar{b}})}{8\pi k}\left[ 2\frac{m}{k}+\left( 1-\frac{4m^{2}}{k^{2}}%
\right) \arctan \frac{k}{2m}\right] +\frac{e^{2}}{4\pi }\varepsilon _{\bar{a}%
\bar{b}\bar{c}}k^{\bar{c}}\left( 1-2\frac{m}{k}\arctan \frac{k}{2m}\right) .
\label{vippa}
\end{equation}
The last term of (\ref{vippa}) contains a local contribution, which survives
in the massless limit and is known as parity anomaly \cite{richter}. Its
sign depends on the sign of the mass, which here was taken to be positive.
On the other hand, this local term is trivial in perturbation theory,
because it can be reabsorbed with a local counterterm, proportional to the
Chern-Simons action of the gauge field. For non-perturbative aspects related
to this issue, especially in non-Abelian gauge theories, the reader is
referred to the literature \cite{richter,parityanom}.

\bigskip

\textbf{One-loop fermion self-energy}. The electron self-energy is an
interesting diagram because both the gauge-field and the fermion propagators
participate. The diagram constructed with the second vertex of (\ref
{vertices}) does not contribute, because it is a massless tadpole. The other
diagram can be computed as follows. First, use Feynman parameters to have
one denominator. Secondly, integrate over $\overline{p}$. This is done
analytically continuing in $D$ and then setting $D=3$. No spurious pole in $%
D-3$ appears. The third step is the integral over the evanescent components $%
t=\hat{p}^{2}$ of the loop momentum. The $t$ integral has the structure of (%
\ref{abdalla3}) with $g=3$ and can be evaluated using formula (\ref{abdalla2}%
). Forth, take the limit $\varepsilon \rightarrow 0$ and fifth, integrate
the result over $x$.

The result is 
\begin{eqnarray*}
SE &=&\frac{i\alpha k\!\!\!\slash}{24\pi }\left[ 2+3\frac{m^{2}}{k^{2}}-%
\frac{3}{k}(ik\!\!\!\slash+m)\left( 1+\frac{m^{2}}{k^{2}}\right) \arctan 
\frac{k}{m}\right] + \\
&&+\frac{1}{8\pi \lambda }\left[ 1+\frac{ik\!\!\!\slash}{2k^{3}}(ik\!\!\!%
\slash+m)^{2}\arctan \frac{k}{m}-\frac{imk\!\!\!\slash}{2k^{2}}\right]
\end{eqnarray*}
and can be checked with an ordinary cut-off method (which however produces
also a linear divergence).

\section{Large N expansion}

\setcounter{equation}{0}

In this section I study the deformed regularization of certain
three-dimensional fermion and scalar models in the large N expansion.

\bigskip

3$D$ \textbf{four-fermion} \textbf{models in the large }$N$\textbf{\
expansion}. The four-fermion theory in three dimensions is described by the
lagrangian 
\[
\mathcal{L}=\overline{\psi }(\partial \!\!\!\slash+m)\psi +\frac{1}{2}%
M\sigma ^{2}+\lambda \sigma \overline{\psi }\psi , 
\]
where $\lambda $ and $M$ are parameters and $\sigma $ is an auxiliary field.
This theory, despite its non-renormalizability by power-counting, can be
defined in the large $N$ expansion \cite{parisi}, resumming the fermion
bubbles (one-loop $\sigma $ self-energy) into an effective $\sigma $
propagator. The ordinary dimensional continuation does not regularize
completely (even if the spinors are four-component), because the resummation
of fermion bubbles gives an effective $\sigma $ propagator of the form 
\begin{equation}
\frac{1}{\mu ^{\varepsilon }(k^{2})^{(1-\varepsilon )/2}+M}  \label{ort}
\end{equation}
that produces $\Gamma (0)$s in subleading Feynman diagrams. Two ways to
circumvent this difficulty have been used in ref.s \cite
{largeN,largeN2,ineq2}: a non-local improvement of the dimensional
technique, valid only with four-component spinors, and a higher-derivative
regularization. Here I consider a more general framework. Recall that the
purpose of this paper is to work out an efficient regularization technique
that is also universal, and in particular admits a straightforward extension
to curved space and non-Abelian gauge theories. In curved space it is
extremely heavy to deal with non-local regularizations, containing
evanescent powers of derivative operators. Moreover, when the spinors are
two-component the usual dimensional continuation of the Dirac algebra is
inconsistent.

It is possible to avoid all this choosing the regularization 
\begin{equation}
\mathcal{L}=\overline{\psi }\left( \overline{\partial }\!\!\!\slash+m-\frac{%
\widehat{\partial ^{2}}}{\Lambda }\right) \psi +\frac{1}{2}\sigma \left( M-%
\frac{\overline{\partial ^{2}}}{\Lambda }+\frac{\widehat{\partial ^{2}}^{2}}{%
\Lambda ^{3}}\right) \sigma +\lambda \sigma \overline{\psi }\psi .
\label{bordo}
\end{equation}
To keep the notation to a minimum, I do not make an explicit distinction
between bare and renormalized quantities. The lagrangian (\ref{bordo}) can
be read either as the bare lagrangian or as the renormalized lagrangian (up
to evanescent counterterms: see (\ref{tora})). In the latter case, it is
understood that appropriate renormalization constants multiply fields and
parameters, and a factor $\mu ^{\varepsilon /2}$ multiplies the vertex $%
\lambda \sigma \overline{\psi }\psi $. Recall that $Z_{\lambda }=1$.

The first interesting quantity to consider in this model is the $\sigma $
self-energy. It is necessary to calculate this diagram for generic values of 
$\varepsilon ,\Lambda $ and the external momentum $(\overline{k},\widehat{k}%
) $. I report here only the result, giving details of the calculation in the
simpler scalar model studied below: 
\begin{equation}
B_{f}(k,m)=\frac{\lambda ^{2}N\mu ^{\varepsilon }}{4\pi \overline{k}}%
2^{3\varepsilon /2}\pi ^{\varepsilon /2}\Lambda ^{-\varepsilon /2}\Gamma
(-2+\varepsilon /2)\ u^{-\varepsilon /4}\left[ \Upsilon \sin \left( \frac{%
\varepsilon }{2}\psi \right) +\Psi \cos \left( \frac{\varepsilon }{2}\psi
\right) \right] ,  \label{abba}
\end{equation}
where 
\begin{eqnarray}
\psi &=&\arctan \frac{\overline{k}}{2\widetilde{m}}\text{ },\text{\qquad
\qquad }u=\overline{k}^{2}+4\widetilde{m}^{2},\qquad \widetilde{m}=m+\frac{%
\widehat{k}^{2}}{4\Lambda },  \label{nona} \\
\Upsilon &=&-(2-\varepsilon )\overline{k}^{2}-8\widetilde{m}^{2},\qquad \Psi
=2\varepsilon \overline{k}\widetilde{m}.  \nonumber
\end{eqnarray}
The spinor is assumed to be two-component. Observe that the evanescent
components $k_{\hat{\mu}}$ of the momentum appear only inside $\widetilde{m}$%
, so $\widehat{k}^{2}/\Lambda $ plays the role of a mass. The $\varepsilon
\rightarrow 0,\Lambda \rightarrow \infty $ limit is convergent, 
\begin{equation}
B_{f}(k,m)\rightarrow -\frac{\lambda ^{2}N}{4\pi \overline{k}}\left[ (%
\overline{k}^{2}+4m^{2})\arctan \frac{\overline{k}}{2m}\right] +\frac{%
\lambda ^{2}N}{2\pi }m.  \label{bolla}
\end{equation}
In the usual cut-off approach \cite{rosen} a linear divergence is generated,
which is cancelled by means of a fine-tuning. The result (\ref{bolla})
agrees with the known one up to a scheme change, because the last term can
be reabsorbed with a redefinition of $M$. Observe that any $\widehat{k}$
dependence has disappeared in the limit.

\bigskip

The effective $\sigma $ propagator $\Sigma (k,m)$ is obtained resumming the
geometric series of the one-loop $\sigma $ self-energies: 
\begin{equation}
\Sigma (k,m)=\frac{1}{M-B_{f}(k,m)+\overline{k}^{2}/\Lambda +\widehat{k}%
^{4}/\Lambda ^{3}}.  \label{effe}
\end{equation}
Now I prove that the term $\overline{k}^{2}/\Lambda +\widehat{k}^{4}/\Lambda
^{3}$ corrects the UV behavior of $B_{f}(k,m)$ and regularizes the $\Gamma
[0]$s in subleading diagrams, avoiding the problems of the naive propagator (%
\ref{ort}). See also \cite{largeN,largeN2,ineq2} on this issue. The proof is
an adaptation of the argument that leads to (\ref{sepia}) and (\ref{sepia2}).

Consider a generic Feynman diagram. The integrand is a polynomial $Q$ in
momenta, due to the vertices and propagators other than (\ref{effe}), times
a certain number $n$ of propagators (\ref{effe}). Consider the $\widehat{k}$%
-integrations. For $\widehat{k}$ large at fixed $\overline{k}$, $%
B_{f}(k,m)\sim $ $\widehat{k}^{2-\varepsilon }/\Lambda $ , so if $\Re
e~\varepsilon >-2$%
\begin{equation}
f(\overline{k})\equiv \int \frac{\mathrm{d}^{-\varepsilon }\widehat{k}}{%
(2\pi )^{-\varepsilon }}[\Sigma (k,m)]^{n}Q(k)\sim \int \frac{\mathrm{d}%
^{-\varepsilon }\widehat{k}}{(2\pi )^{-\varepsilon }}\sum_{p,q}\frac{%
\widehat{k}^{-\varepsilon q}}{\widehat{k}^{p}}\sim \sum_{p,q}\frac{\Gamma
\left( p/2+(q+1)\varepsilon /2\right) }{\Gamma (p/2+q\varepsilon /2)},
\label{ossi}
\end{equation}
where $q$ and $p$ are integers, $q\geq 0$. Consequently, no $\Gamma [0]$ is
generated in the numerator provided that $\Re e~\varepsilon >-2$. The
requirement $\Re e~\varepsilon >-2$ is compatible with the usual conditions
for the existence of convergence domains for the integrals. Indeed, these
conditions have the form $\delta _{\mathrm{UV}}<\Re e~\varepsilon <\delta _{%
\mathrm{IR}}$, for some integers $\delta _{\mathrm{IR}}\geq 0$ and $\delta _{%
\mathrm{UV}}<\delta _{\mathrm{IR}}$ (see Appendix A for more details). The
inequality $\delta _{\mathrm{IR}}\geq 0$ is due to the fact that the
integrands are regular for $\widehat{k}\rightarrow 0$. The subsets $\Re
e~\varepsilon >-2$ and $\delta _{\mathrm{UV}}<\Re e~\varepsilon <\delta _{%
\mathrm{IR}}$ have always a non-empty intersection.

Now consider the $\overline{k}$-integration 
\[
\int \frac{\mathrm{d}^{D}\overline{k}}{(2\pi )^{D}}f(\overline{k}) 
\]
on the assumption that the $\widehat{k}$-integration has already been done.
To study the large-$\overline{k}$ behavior of $f(\overline{k})$, rescale $%
\overline{k}$ by a factor $\lambda $ in $f(\overline{k})$. At the same time,
rescale $\widehat{k}$ by a factor $\sqrt{\lambda }$ in the $\widehat{k}$%
-integral that defines $f(\overline{k})$, see (\ref{ossi}). Observing that $%
B_{f}(k,m)$ goes into $\lambda ^{1-\varepsilon /2}B_{f}(k,m/\lambda )$ and
repeating the argument used for (\ref{ossi}), we find, for $\Re
e~\varepsilon >-2$, 
\[
f(\lambda \overline{k})\sim \sum_{p,q}\lambda ^{-p-(q+1)\varepsilon /2}\text{%
,\qquad therefore }\int \frac{\mathrm{d}^{D}\overline{k}}{(2\pi )^{D}}f(%
\overline{k})\sim \sum_{p,q}\frac{\Gamma \left( p/2+(q+1)\varepsilon
/4-D/2\right) }{\Gamma \left( p/2+(q+1)\varepsilon /4\right) }, 
\]
where $q$ and $p$ are integers, $q\geq 0$. Again, no $\Gamma [0]$ is
generated.

\bigskip

3$D$ \textbf{scalar conformal field theories} \textbf{in the large }$N$%
\textbf{\ expansion}. The three-dimensional scalar model 
\begin{equation}
\mathcal{L}_{\text{scalar}}=\frac{1}{2}\sum_{i=1}^{N}\left[ \left( \partial
_{\mu }\varphi _{i}\right) ^{2}+i\lambda \sigma \varphi _{i}^{2}\right] ,
\label{scala}
\end{equation}
where $\sigma $ is a dynamical field (see \cite{largeN2}), describes a
conformal field theory, the UV (Wilson-Fischer) fixed point of the $O(N)$
sigma model. Although this theory can be regularized also in a standard way,
it is instructive to describe how to proceed in the deformed framework. For
massive scalars the regularized lagrangian reads 
\[
\mathcal{L}_{\text{scalar}}=\frac{1}{2}\sum_{i=1}^{N}\left[ \left( \overline{%
\partial }_{\mu }\varphi _{i}\right) ^{2}+\left( \frac{\widehat{\partial ^{2}%
}}{\Lambda }\varphi _{i}-m\varphi _{i}\right) ^{2}+i\lambda \sigma \varphi
_{i}^{2}\right] +\frac{1}{2\Lambda }\sigma ^{2}. 
\]
First I describe the calculation of the scalar bubble (one-loop $\sigma $
self-energy) for $m=0$, then I add the mass. 
\begin{figure}[tbp]
\begin{center}
\epsfig{figure=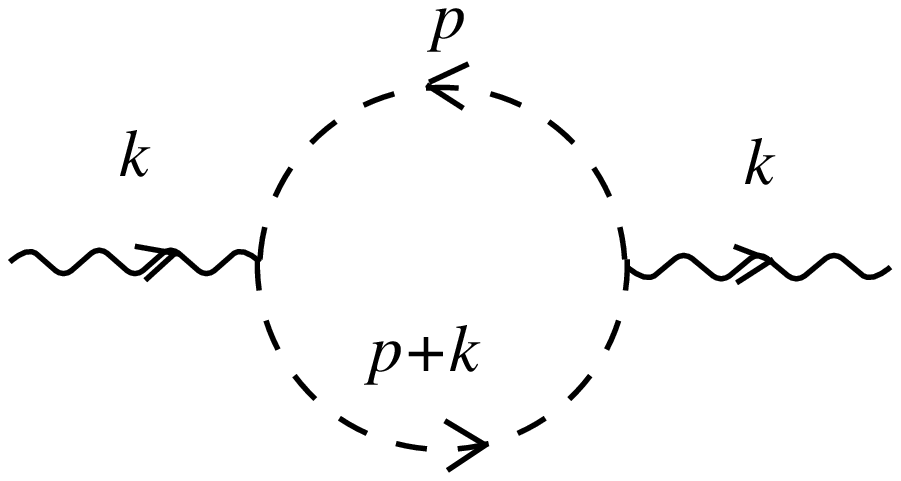,height=2cm,width=7cm}
\end{center}
\end{figure}
The integral over the physical components $\overline{p}$ of the loop
momentum can be done easily, and gives 
\begin{equation}
-\frac{i\lambda ^{2}N\mu ^{\varepsilon }}{16\pi \overline{k}}\ln \frac{2t+%
\widehat{k}^{2}-i\overline{k}\Lambda +2\widehat{p}\cdot \widehat{k}}{2t+%
\widehat{k}^{2}+i\overline{k}\Lambda +2\hat{p}\cdot \widehat{k}}.
\label{resa}
\end{equation}
The next task is the integral over the angle between $\widehat{p}$ and $%
\widehat{k}$. This is done expanding (\ref{resa}) in powers of $\widehat{p}%
\cdot \widehat{k}$. By symmetric integration, it is easy to show that the
angular integration of a power $\left( \widehat{p}\cdot \widehat{k}\right)
^{n}$, multiplied by a function of $\widehat{p}^{2}$ and $\widehat{k}^{2}$,
is equivalent to the substitution 
\[
\left( \widehat{p}\cdot \widehat{k}\right) ^{n}\rightarrow \frac{\Gamma
(n/2+1/2)\Gamma (-\varepsilon /2)}{\sqrt{\pi }\Gamma (n/2-\varepsilon /2)}%
\left( \widehat{p}^{2}\widehat{k}^{2}\right) ^{n/2}\text{ if }n\text{ is
even,} 
\]
while it gives $0$ if $n$ is odd. At this point the integral over $\widehat{p%
}^{2}$ is done term-by-term in the expansion. Finally, the series in $n$ is
resummed. The result is 
\begin{equation}
B_{s}(k)=-\frac{\lambda ^{2}N\mu ^{\varepsilon }}{8\pi \overline{k}}%
2^{2\varepsilon }\pi ^{\varepsilon /2}\Gamma (\varepsilon /2)\left( \widehat{%
k}^{4}+4\overline{k}^{2}\Lambda ^{2}\right) ^{-\varepsilon /4}\sin \left( 
\frac{\varepsilon }{2}\arctan \frac{2\overline{k}\Lambda }{\widehat{k}^{2}}%
\right) .  \label{renb}
\end{equation}

The calculation can be extended to massive scalars. The intermediate result (%
\ref{resa}) is modified adding $2m\Lambda $ both to the numerator and
denominator of the fraction inside the logarithm. Finally, with a simple
replacement, the generalization of (\ref{renb}) is 
\begin{equation}
B_{s}(k,m)=-\frac{N\lambda ^{2}\mu ^{\varepsilon }2^{3\varepsilon /2}\pi
^{\varepsilon /2}\Lambda ^{-\varepsilon /2}}{8\pi \overline{k}}\Gamma
(\varepsilon /2)\ u^{-\varepsilon /4}\sin \left( \frac{\varepsilon }{2}\psi
\right) ,  \label{scal}
\end{equation}
where $u$ and $\psi $ are defined as in (\ref{nona}).

The $\sigma $ self-energies can be resummed into the effective $\sigma $%
-propagator, which can be used to calculate the $\mathcal{O}(1/N)$
subleading corrections to anomalous dimensions and other quantities \cite
{largeN2}. Despite the complicated structure of (\ref{scal}), the
high-energy behavior of $B_{s}$ is simpler, and sufficient to calculate the
divergent parts of Feynman diagrams. It is also not difficult to calculate
the beta functions of the RG flows of \cite{largeN2} that interpolate in a
classically conformal way between fixed points of the type (\ref{scala}).

With the same procedure and some more algebra it is possible to derive the
result (\ref{abba}).

Summarizing, it is possible to evaluate complete amplitudes such as $%
B_{f}(k,m)$ and $B_{s}(k,m)$, that contain two arbitrary cut-offs and depend
on a mass and generic physical and evanescent momenta. This is another
indication that the regularization defined here can be used efficiently.

\section{Evanescent higher-derivative deformation in flat space}

\setcounter{equation}{0}

Several fields admit an $SO(d)$ invariant dimensional regularization. Other
fields do not. If a theory contains fields of both types it might be
convenient to deform also the regularization of the $SO(d)$ invariant
fields, so that all propagators have denominators with the same structure
(in the massless limit). This makes the use of Feynman parameters more
efficient.

\bigskip

\textbf{Scalar fields}. For scalar fields, the lagrangian 
\begin{equation}
\mathcal{L}_{\text{scalar}}=\frac{1}{2}\left[ \left( \overline{D}_{\mu
}\varphi \right) ^{2}+\left( \frac{\widehat{D^{2}}}{\Lambda }\varphi
-m\varphi \right) ^{2}\right]  \label{scaldue}
\end{equation}
produces denominators with the same structure as for fermions also at $m\neq
0$.

\bigskip

\textbf{Gauge vectors}. In (Abelian and non-Abelian) Yang-Mills theory, the
lagrangian 
\begin{eqnarray}
\mathcal{L} &=&\mathcal{L}_{\text{vector}}+\mathcal{L}_{\text{gf}}+\mathcal{L%
}_{\text{ghost}},\qquad \mathcal{L}_{\text{vector}}=\frac{1}{4\alpha }\left(
F_{\bar{\mu}\bar{\nu}}^{2}-2F_{\bar{\mu}\hat{\nu}}\frac{\widehat{D^{2}}}{%
\Lambda ^{2}}F_{\bar{\mu}\hat{\nu}}+F_{\hat{\mu}\hat{\nu}}\frac{\widehat{%
D^{2}}^{2}}{\Lambda ^{4}}F_{\hat{\mu}\hat{\nu}}\right) ,  \label{qed} \\
\mathcal{L}_{\text{gf}} &=&\lambda \left( \overline{\partial A}-\frac{1}{%
\Lambda ^{2}}\widehat{\partial ^{2}}\widehat{\partial A}\right) ^{2},\qquad
\qquad \mathcal{L}_{\text{ghost}}=\overline{C}\left( -\overline{\partial D}+%
\widehat{\partial ^{2}}\frac{\widehat{\partial D}}{\Lambda ^{2}}\right) C,
\label{qed2}
\end{eqnarray}
where $D_{\mu }$ denotes the covariant derivative, while $D^{2}=D_{\mu
}D^{\mu }$, $\partial D=\partial _{\mu }D^{\mu }$ etc., produces the
propagators 
\begin{eqnarray}
\!\!\!\!\!\!\!\!\!\!\!\!\langle A_{\mu }(p)~A_{\nu }(-p)\rangle _{\mathrm{%
free}} &=&\frac{\alpha }{\overline{p}^{2}+\widehat{p}^{4}/\Lambda ^{2}}\left[
\overline{\delta }_{\mu \nu }+\frac{\Lambda ^{2}}{\widehat{p}^{2}}\widehat{%
\delta }_{\mu \nu }+\left( \frac{1}{2\lambda \alpha }-1\right) \frac{p_{\mu
}p_{\nu }}{\overline{p}^{2}+\widehat{p}^{4}/\Lambda ^{2}}\right] ,  \nonumber
\\
\langle C(p)~\overline{C}(-p)\rangle _{\mathrm{free}} &=&\frac{1}{\overline{p%
}^{2}+\widehat{p}^{4}/\Lambda ^{2}}.  \label{barba}
\end{eqnarray}
Again, the structure of denominators simplifies the use of Feynman
parameters in the $\overline{p}$ integration. However, there appears a
denominator $\widehat{p}^{2}$, which can be responsible of IR divergences.
To avoid this nuisance it is safer to further deform with an evanescent
mass. This is achieved replacing $\widehat{D^{2}}$ and $\widehat{\partial
^{2}}$ in (\ref{qed}-\ref{qed2}) with $\widehat{D^{2}}-\widehat{m}^{2}$ and $%
\widehat{\partial ^{2}}-\widehat{m}^{2}$, respectively.

\bigskip

There exists a simple recipe to construct a manifestly gauge invariant
higher-derivative deformation of an $SO(d)$ invariant lagrangian in flat
space. Start from the $SO(d)$ invariant lagrangian and perform the
replacements 
\begin{equation}
A_{\mu }\rightarrow A_{\mu },\qquad \partial _{\mu }\rightarrow \partial
_{\mu },\qquad \delta ^{\mu \nu }\rightarrow \delta ^{\bar{\mu}\bar{\nu}}+%
\frac{\widehat{m}^{2}-\widehat{D^{2}}}{\Lambda ^{2}}\delta ^{\hat{\mu}\hat{%
\nu}},\qquad \delta _{\nu }^{\mu }\rightarrow \delta _{\nu }^{\mu }\text{.}
\label{repla}
\end{equation}
Observe that lower and upper indices have to be kept distinct during the
replacement. The rules (\ref{repla}) imply also 
\begin{eqnarray}
F_{\mu \nu } &\rightarrow &F_{\mu \nu },\text{\qquad }D_{\mu }=\partial
_{\mu }+iA_{\mu }\rightarrow D_{\mu },\qquad \delta _{\mu \nu }\rightarrow
\delta _{\bar{\mu}\bar{\nu}}+\frac{\Lambda ^{2}}{\widehat{m}^{2}-\widehat{%
D^{2}}}\delta _{\hat{\mu}\hat{\nu}},  \nonumber \\
D^{\mu } &\rightarrow &D^{\bar{\mu}}+\frac{\widehat{m}^{2}-\widehat{D^{2}}}{%
\Lambda ^{2}}D^{\hat{\mu}},\qquad D^{2}\rightarrow \overline{D^{2}}+\frac{%
\left( \widehat{m}^{2}-\widehat{D^{2}}\right) }{\Lambda ^{2}}\widehat{D^{2}}.
\label{repla2}
\end{eqnarray}
The transformation rules of $A^{\mu }$, $F^{\mu \nu }$ etc., follow
consequently. Gauge invariance is manifest. The replacement is local, namely
the deformation of a local lagrangian is a local lagrangian. Indeed, only
the replacement of $\delta _{\mu \nu }$ contains a non-local term, but this
affects the propagator, not the lagrangian.

Since $\delta ^{\mu \nu }$ is deformed into a derivative operator, is is
necessary to specify the position of the tensor $\delta ^{\mu \nu }$ before
the replacement. The position of $\delta ^{\mu \nu }$ is determined
observing that the quadratic part of the lagrangian is correctly deformed
only if $\delta ^{\mu \nu }$ is placed in between the fields (the
deformation would have no effect otherwise). For example, if $\Phi _{\mu }$
is a vector field, 
\[
\Phi ^{2}\equiv \Phi _{\mu }\delta ^{\mu \nu }\Phi _{\nu }\rightarrow 
\overline{\Phi }^{2}+\widehat{\Phi }\frac{\widehat{m}^{2}-\widehat{D^{2}}}{%
\Lambda ^{2}}\widehat{\Phi }. 
\]
After the replacement there is no need to distinguish between upper and
lower indices.

At $\widehat{m}=0$ the replacement (\ref{repla}) produces immediately the
quadratic part of (\ref{qed}-\ref{qed2}) and the propagators (\ref{barba}).
In the gauge-fixing and ghost terms (\ref{qed2}) the covariant box $\widehat{%
D^{2}}$ can be replaced with the simple box $\widehat{\partial^2 }$. As a
consistency check, observe that when $\Lambda ^{2}$ is (formally) set equal
to $-\widehat{D^{2}}$ or $-\widehat{\partial^2 }$ in (\ref{qed}-\ref{qed2})
and (\ref{barba}) the standard $SO(d)$ invariant lagrangian as well as the $%
SO(d)$ invariant propagators are obtained. Moreover, the replacement (\ref
{repla}) becomes the identity in this formal limit.

At $\widehat{m}\neq 0$ the evanescent mass $\widehat{m}$ takes care of the
IR\ nuisances mentioned above, due to the denominator $1/\widehat{p}^{2}$ in
(\ref{barba}). Observe that at $\widehat{m}\neq 0$ the calculations done so
far do not become conceptually more difficult than at $\widehat{m}=0$.

As far as fermions are concerned, the action (\ref{a1}) is produced with the
additional rule 
\begin{equation}
\gamma ^{\mu }\rightarrow \gamma ^{\bar{\mu}}-\frac{1}{\Lambda }D^{\hat{\mu}}
\label{gamma}
\end{equation}
so that 
\[
D\!\!\!\!\slash=\gamma ^{\mu }D_{\mu }\rightarrow \left( \gamma ^{\bar{\mu}}-%
\frac{1}{\Lambda }D^{\hat{\mu}}\right) D_{\mu }=\overline{D\!\!\!\!\slash}-%
\frac{\widehat{D^{2}}}{\Lambda }. 
\]

With Chern-Simons gauge fields it is possible to proceed as follows.
Inspired by the replacement (\ref{gamma}), write 
\begin{equation}
\varepsilon ^{\mu \rho \nu }\partial _{\rho }=-\frac{i}{2}\mathrm{tr}[\gamma
^{\mu }\overline{\partial \!\!\!\slash}\gamma ^{\nu }]\rightarrow -\frac{i}{2%
}\mathrm{tr}\left[ \left( \gamma ^{\bar{\mu}}-\frac{1}{\Lambda }\partial ^{%
\hat{\mu}}\right) \left( \overline{\partial \!\!\!\slash}-\frac{\widehat{%
\partial ^{2}}}{\Lambda }\right) \left( \gamma ^{\bar{\nu}}-\frac{1}{\Lambda 
}\partial ^{\hat{\nu}}\right) \right] .  \label{inte}
\end{equation}
This replacement is however incomplete, because it is not gauge invariant,
namely it is not annihilated by the contractions with $\partial _{\nu }$ and 
$\partial _{\mu }$. It is possible to complete it changing a sign in (\ref
{inte}) and adding a term: 
\begin{equation}
\varepsilon ^{\mu \rho \nu }\partial _{\rho }\rightarrow -\frac{i}{2}\mathrm{%
tr}\left[ \left( \gamma ^{\bar{\mu}}-\frac{1}{\Lambda }\partial ^{\hat{\mu}%
}\right) \left( \overline{\partial \!\!\!\slash}+\frac{\widehat{\partial ^{2}%
}}{\Lambda }\right) \left( \gamma ^{\bar{\nu}}-\frac{1}{\Lambda }\partial ^{%
\hat{\nu}}\right) \right] -\frac{i}{\Lambda }\delta ^{\hat{\mu}\hat{\nu}%
}\left( \overline{\partial ^{2}}-\frac{\widehat{\partial ^{2}}^{2}}{\Lambda
^{2}}\right) .  \label{inte2}
\end{equation}
Expanding and reorganizing, the expression (\ref{lchs}) is immediately
recovered.

\section{Absence of power-like divergences}

\setcounter{equation}{0}

The second cut-off $\Lambda $ appears explicitly in the regularized
lagrangian, so (gauge invariant) power-like divergences can in principle be
generated. This does happen if regularized lagrangians such as (\ref{sire})
and (\ref{luno}) are used. However, when the higher-derivative deformation
is also evanescent, as in (\ref{a1}), (\ref{lchs}), (\ref{scaldue}) and (\ref
{qed}-\ref{qed2}), then the power-like divergences are set to zero by
default. In this section I\ prove this statement, confirmed by the results
of the calculations of the previous sections, and derive other properties of
the counterterms.

Consider first the regularized Dirac action (\ref{a1}). If the evanescent
components $A_{\hat{\mu}}$ of the gauge vector and the evanescent spacetime
coordinates $x^{\hat{\mu}}$ are rescaled as follows 
\begin{equation}
A_{\hat{\mu}}=\widetilde{A}_{\hat{\mu}}\sqrt{\Lambda },\qquad x^{\hat{\mu}}=%
\widetilde{x}^{\hat{\mu}}/\sqrt{\Lambda },  \label{rescala}
\end{equation}
then the $\Lambda $ dependence reduces just to a factor $\Lambda
^{\varepsilon /2}$ in front of the lagrangian. The same holds for the
regularized Chern-Simons lagrangian (\ref{lchs}-\ref{borabora}) and the
deformed Yang-Mills lagrangian (\ref{qed}-\ref{qed2}). At $\widehat{m}\neq 0$
it is necessary to rescale also $\widehat{m}$ to $\widehat{\widetilde{m}}%
\sqrt{\Lambda }$. Correspondingly, in the Feynman diagrams a rescaling of
the integrated momenta, $\widehat{p}\rightarrow \widehat{\widetilde{p}}\sqrt{%
\Lambda }$, factorizes a certain power of $\Lambda ^{\varepsilon /2}$, so
the divergent parts have the form 
\begin{equation}
\frac{\Lambda ^{q\varepsilon /2}}{\varepsilon ^{n}}  \label{arra}
\end{equation}
times a function of external momenta (with components $\overline{k},\widehat{%
\widetilde{k}}$), $\widehat{\widetilde{m}}$, etc., where $q$ and $n$ are
integers. Then, replacing the tilded-hatted objects with the original hatted
ones, only negative powers of $\Lambda $ are produced, but no positive
power. On the other hand, it is evident that the expansion of (\ref{arra})
in powers of $\varepsilon $ produces only logarithms of $\Lambda $. This
proves that no power-like divergences are generated.

After the rescaling (\ref{rescala}), the evanescent components of fields,
coordinates and momenta have non-canonical dimensionalities. For example, in 
$D=3$ the Chern-Simons gauge field $A_{\mu }=(A_{\bar{\mu}},A_{\hat{\mu}})$
has, before the rescaling, dimensionality $1$. After the rescaling (\ref
{rescala}) the physical components $A_{\bar{\mu}}$ keep their dimensionality 
$1$, but the tilded evanescent components $\widetilde{A}_{\hat{\mu}}$
acquire dimensionality $1/2$. Similarly, $p_{\bar{\mu}}$ has dimensionality
1, but $\widetilde{p}_{\hat{\mu}}$ has dimensionality 1/2. It is easy to
check that the dimensionalities of the physical and tilded-evanescent
components of fields and momenta are always strictly positive. By the
theorem of locality of the counterterms, the counterterms are polynomial in
these quantities, and therefore polynomial also in the untilded quantities.
In particular, the poles in $\varepsilon $ cannot multiply arbitrary
negative powers of $\Lambda $.

Summarizing, at $\Lambda $ fixed the poles in $\varepsilon $ can multiply
logarithms of $\Lambda $, a finite number of negative powers of $\Lambda $,
and no positive power of $\Lambda $.

The argument just outlined does not work if different fields are regularized
in an incoherent way. For example, consider two-component fermions coupled
with Yang-Mills theory in three dimensions. The fermions cannot be
regularized in a $SO(d)$ invariant way, but can be regularized as in (\ref
{a1}). If the gauge fields are regularized in a $SO(d)$ invariant way, the
rescaling (\ref{rescala}) does not reduce the $\Lambda $ dependence of the
complete lagrangian to just a factor $\Lambda ^{\varepsilon /2}$ in front of
it. When the theory contains some fields that do not admit an $SO(d)$
invariant regularization, it is convenient to deform also the fields that do
admit one, using the replacement (\ref{repla}-\ref{repla2}). Then the
argument based on the rescaling (\ref{rescala}) applies as described above.

The evanescent deformation (\ref{repla}-\ref{repla2}) always exists in flat
space and it is immediate to generalize it to curved space, as long as
gravity is not quantized. Instead, I have no simple generalization of the
evanescent deformation to quantum gravity. The difficulty is to find a
suitable higher-derivative deformation of the Einstein lagrangian. Combining
the regularization of this paper with the one of ref. \cite{pap4SM} it is
possible to dimensionally regularize also odd-dimensional parity-violating
theories coupled with quantum gravity in a manifestly gauge-invariant way,
for example the models of \cite{pap1,pap2}. However, mixed ($SO(d)$
invariant and $SO(d)$ non-invariant) denominators appear in Feynman diagrams
and it might be necessary to eliminate power-like divergences manually.

\section{Conclusions}

\setcounter{equation}{0}

The results of this paper suggest that there always exists an appropriate
deformation of the dimensional-regularization technique that regularizes
consistently and in a manifestly gauge-invariant way also the models to
which the naive dimensional technique does not apply. In the deformed
framework, the spacetime dimension is still analytically continued to
complex values, but manifest Lorentz invariance is restricted to the
physical subsector of spacetime. Then it is possible to use the ordinary
(uncontinued) Dirac algebra, gaining a certain simplification of the
renormalization structure. The regularization is completed with an
evanescent higher-derivative deformation, which makes use of an extra
cut-off. At higher orders, evanescent counterterms can be subtracted just as
they come, without spoiling the convenient tree-level structure of the
regularized lagrangian.

The virtues of the deformed regularization are that it is universal, local,
manifestly gauge invariant (up to the known anomalies) and Lorentz invariant
in the physical sector of spacetime. In flat space it kills power-like
divergences by default. Infinitely many evanescent operators are
automatically dropped. I have paid special attention to the efficiency of
practical computations.

The existence of a universal regularization technique with the properties
just mentioned is useful to quickly prove the absence of gauge anomalies in
the models where the ordinary dimensional technique is inconsistent, in
particular when composite operators of high dimensionalities are considered
or the theory contains non-renormalizable interactions. Alternative proofs
of the absence of gauge anomalies are provided by the
algebraic-renormalization approach \cite{qualcuno}, which does not need an
explicit regularization framework and makes use of an involved cohomological
classification. Another popular framework uses the
exact-renormalization-group techniques \cite{parmensi}, but heavy cut-off
dependencies are generated and Slavnov-Taylor identities are imposed
step-by-step.

The technique of this paper has applications to the study of finiteness and
renormalizability beyond power-counting \cite{pap3,pap2}, but can be
convenient also in four-dimensional renormalizable theories, to reduce the
number of evanescent counterterms.

\vskip 25truept \noindent {\Large \textbf{Acknowledgements}} \vskip 15truept

\noindent I would like to thank the referee for drawing my attention to an
inaccuracy in the first version of the paper.

\vskip 25truept \noindent {\Large \textbf{A\ \ Appendix: useful integrals}} %
\vskip 15truept

\renewcommand{\theequation}{A.\arabic{equation}} \setcounter{equation}{0}

\noindent In this appendix I collect some useful integrals. Let $%
d=D-\varepsilon $, where $D$ denotes the physical spacetime dimension and $d$
denotes the continued spacetime dimension.

In dimensional regularization, an integral $I(\varepsilon )$ is said to
admit a convergence domain if there exists an open set $\mathcal{D}_{I}$ of
the complex plane, such that $I(\varepsilon )$ is convergent for $%
\varepsilon \in \mathcal{D}_{I}$. If an integral $I(\varepsilon )$ admits a
convergence domain $\mathcal{D}_{I}$, then it is first evaluated for $%
\varepsilon \in \mathcal{D}_{I}$ and later extended to the complex plane (up
to eventual poles) by analytical continuation. An integral that admits no
convergence domain can be calculated if it can be split into a finite sum of
integrals that separately admit convergence domains. For the types of
integrals that appear in perturbative quantum field theory, these operations
are consistent and unambiguous.

The situation where the integral does not admit a convergence domain is
frequent in dimensional regularization. For example, the integral of 1 does
not admit a convergence domain, but can be calculated writing it as the sum
of two integrals that separately admit convergence domains: 
\[
\int \frac{\mathrm{d}^{d}p}{(2\pi )^{d}}1=\int \frac{\mathrm{d}^{d}p}{(2\pi
)^{d}}\frac{m^{2}}{p^{2}+m^{2}}+\int \frac{\mathrm{d}^{d}p}{(2\pi )^{d}}%
\frac{p^{2}}{p^{2}+m^{2}}=\frac{\Gamma (1-d/2)}{(4\pi )^{d/2}}m^{d}+\frac{d}{%
2}\frac{\Gamma (-d/2)}{(4\pi )^{d/2}}m^{d}=0. 
\]
The first integral is convergent for $0<\Re e~d<2$, while the second
integral is convergent for $-2<\Re e~d<0$.

An integral frequently met in the paper is 
\begin{equation}
I_{1}=\int_{0}^{\infty }t^{-1-\varepsilon /2}~\mathrm{d}t~\ln
(k_{x}^{2}+(t+m)^{2}).  \label{int}
\end{equation}
There is no complex domain of $\varepsilon $ that makes this integral
convergent. To calculate (\ref{int}), first rewrite it as $2a_{1}+a_{2}$,
where 
\[
a_{1}=\int_{0}^{\infty }t^{-1-\varepsilon /2}~\mathrm{d}t~\ln (t+m),\qquad
a_{2}=\int_{0}^{\infty }t^{-1-\varepsilon /2}~\mathrm{d}t~\ln \frac{%
k_{x}^{2}+(t+m)^{2}}{(t+m)^{2}}. 
\]
The integral $a_{1}$ still does not admit a convergence domain. Multiplying
and dividing the integrand by $t+m$, $a_{1}$ can be split into the sum of
two integrals that separately admit convergence domains: 
\begin{equation}
a_{1}=\int_{0}^{\infty }t^{-\varepsilon /2}~\mathrm{d}t~\frac{\ln (t+m)}{t+m}%
+m\int_{0}^{\infty }t^{-1-\varepsilon /2}~\mathrm{d}t~\frac{\ln (t+m)}{t+m}=%
\frac{2\pi m^{-\varepsilon /2}}{\varepsilon \sin (\pi \varepsilon /2)}.
\label{cory}
\end{equation}
The integral $a_{2}$ admits a convergence domain. It can be safely
calculated expanding the logarithm in powers series as follows:

\begin{equation}
a_{2}=-\sum_{n=1}^{\infty }~\frac{(-1)^{n}}{n}(k_{x}^{2})^{n}\int_{0}^{%
\infty }t^{-1-\varepsilon /2}~\mathrm{d}t~(t+m)^{-2n},  \label{acio}
\end{equation}
then integrating term-by-term and resumming. The result is 
\begin{equation}
a_{2}=\frac{4\pi m^{-\varepsilon /2}}{\varepsilon \sin (\pi \varepsilon /2)}%
\left[ -1+(1+k^{2}x(1-x)/m^{2})^{-\varepsilon /4}\cos \left( \frac{%
\varepsilon }{2}\arctan \frac{k}{m}\sqrt{x(1-x)}\right) \right] .
\label{ayt}
\end{equation}
The total is 
\[
I_{1}=2a_{1}+a_{2}=\frac{4\pi }{\varepsilon \sin (\pi \varepsilon /2)}\left(
k^{2}x(1-x)+m^{2}\right) ^{-\varepsilon /4}\cos \left( \frac{\varepsilon }{2}%
\arctan \frac{k}{m}\sqrt{x(1-x)}\right) . 
\]
It is now safe to expand in powers of $\varepsilon $, obtaining 
\begin{equation}
I_{1}=\frac{8}{\varepsilon ^{2}}-\frac{2}{\varepsilon }\ln
(k^{2}x(1-x)+m^{2})+\mathcal{O}(1).  \label{cicio}
\end{equation}

\bigskip

To evaluate formula (\ref{alla}) it is necessary to calculate the integral 
\begin{equation}
I_{2}=\int_{0}^{\infty }t^{-1-\varepsilon /2}~\mathrm{d}t~\left[
k_{x}^{2}+(t+m)^{2}\right] ^{-1/2}.  \label{appa}
\end{equation}
This can be done with the same procedure as for (\ref{int}). First expand
the integrand in powers of $k_{x}^{2}/(t+m)^{2}$. Then integrate each term
of the expansion over $t$. Finally, resum the power series. The result is 
\begin{equation}
I_{2}=-\frac{\pi m^{-1-\varepsilon /2}}{\sin (\pi \varepsilon /2)}~_{2}F_{1}%
\left[ \frac{1}{2}+\frac{\varepsilon }{4},1+\frac{\varepsilon }{4};1;-\frac{%
k_{x}^{2}}{m^{2}}\right] .  \label{abdalla}
\end{equation}
When $\varepsilon $ tends to zero, the behavior of $I_{2}$ is 
\begin{equation}
I_{2}=-\frac{2}{\varepsilon }\left( k_{x}^{2}+m^{2}\right) ^{-1/2}+\mathcal{O%
}(1),  \label{appa2}
\end{equation}
which, inserted into (\ref{alla}), gives (\ref{vippa}), after a
straightforward integration over $x$.

The result (\ref{abdalla}) can be generalized immediately to give 
\begin{eqnarray}
F_{g} &=&\int_{0}^{\infty }t^{-1-\varepsilon /2}~\mathrm{d}t~\left[
k_{x}^{2}+(t+m)^{2}\right] ^{-g/2}=  \nonumber \\
&=&m^{-g-\varepsilon /2}\frac{\Gamma (g+\varepsilon /2)\Gamma (-\varepsilon
/2)}{\Gamma (g)}~_{2}F_{1}\left[ \frac{g}{2}+\frac{\varepsilon }{4},\frac{g+1%
}{2}+\frac{\varepsilon }{4};\frac{g+1}{2};-\frac{k_{x}^{2}}{m^{2}}\right] .
\label{abdalla2}
\end{eqnarray}
It is also straightforward to calculate integrals of the form 
\begin{equation}
\int_{0}^{\infty }t^{-1-\varepsilon /2}~\mathrm{d}t~P(t)~\left[
k_{x}^{2}+(t+m)^{2}\right] ^{-g/2},  \label{abdalla3}
\end{equation}
$P(t)$ being an arbitrary polynomial in $t$. The result is a sum of terms of
the form (\ref{abdalla2}), with $\varepsilon $ shifted by integer numbers.

Finally, calculating the $-\varepsilon $ integration before the $D$
integration, such as in (\ref{bona}), it is frequent to meet integrals such
as

\begin{equation}
I[p,n]\equiv \int_{0}^{\infty }\mathrm{d}s\frac{s^{p}}{%
\prod_{i=1}^{n}(s+a_{i})}=\frac{\pi (-1)^{n}}{\sin p\pi }\sum_{i=1}^{n}\frac{%
a_{i}^{p}}{\prod_{j\neq i}(a_{i}-a_{j})}.  \label{mil}
\end{equation}
I have checked this formula in various cases (up to $n=4$ included with
different $a$s, for special values of the $a$s with higher $n$). It
satisfies the recursion relation 
\[
I[p,n]=I[p+1,n+1]+a_{n+1}I[p,n+1]. 
\]

\vskip 25truept \noindent {\Large \textbf{B Appendix: non-evanescent
higher-derivative deformations}} \vskip 15truept

\renewcommand{\theequation}{B.\arabic{equation}} \setcounter{equation}{0}

For completeness, in this appendix I collect some alternative
higher-derivative deformations, which are equally consistent, but less
efficient in practical computations.

An obvious alternative to the regularized Dirac action (\ref{a1}) is

\begin{equation}
\mathcal{L}_{\text{Dirac}}=\overline{\psi }\left( \overline{D\!\!\!\!\slash}-%
\frac{D^{a}D_{a}}{\Lambda }\right) \psi .  \label{sire}
\end{equation}
Calculations with the propagator induced by this action are however more
involved. A less obvious alternative is 
\begin{equation}
\mathcal{L}_{\text{Dirac}}=\overline{\psi }\left( \overline{D\!\!\!\!\slash}%
+i\widehat{D\!\!\!\!\slash}\right) \psi .  \label{a2}
\end{equation}
The evanescent correction is imaginary, and Hermiticity (or reflection
positivity, in the Euclidean framework) is retrieved in the limit $%
\varepsilon \rightarrow 0$. Here no additional cut-off $\Lambda $ is needed.
The lagrangian (\ref{a2}) is useful when fermions are massless, because the
propagator 
\begin{equation}
\frac{1}{i\overline{p\!\!\!\slash}-\widehat{p\!\!\!\slash}}=\frac{-i%
\overline{p\!\!\!\slash}-\widehat{p\!\!\!\slash}}{p^{2}}  \label{propafe}
\end{equation}
has an $SO(d)$ invariant denominator. The violation of chiral invariance is
due to $[\gamma _{5},\gamma ^{\hat{a}}]=0$. The computation of the axial
anomaly with this action resembles the usual computation in dimensional
regularization \cite{collins}. However, when the fermions are massive the
propagator is not a simple modification of (\ref{propafe}), but 
\[
\frac{(-i\overline{p\!\!\!\slash}-\widehat{p\!\!\!\slash}+m)(p^{2}+m^{2}+2m%
\widehat{p\!\!\!\slash})}{(p^{2}+m^{2})^{2}-4\widehat{p}^{2}m^{2}}, 
\]
which makes computations rather hard.

The simplest higher-derivative deformation of the Chern-Simons lagrangian (%
\ref{chernsi}) is 
\begin{equation}
\mathcal{L}_{\text{ChS}}=-\frac{i}{2\alpha }\varepsilon _{\bar{a}\bar{b}\bar{%
c}}F^{\bar{a}\bar{b}}A^{\bar{c}}+\frac{1}{4\Lambda }F_{\mu \nu }^{2}.
\label{luno}
\end{equation}
The gauge-fixing term $(\partial _{\mu }A^{\mu })^{2}/(2\lambda )$ produces,
in the Feynman gauge $\lambda =\Lambda $, the propagator 
\begin{equation}
\langle A_{\mu }(p)~A_{\nu }(-p)\rangle _{\mathrm{free}}=\frac{\Lambda }{%
p^{2}}\delta _{\mu \nu }+\frac{\Lambda (p_{\bar{\mu}}p_{\bar{\nu}}-\overline{%
p}^{2}\delta _{\bar{\mu}\bar{\nu}})}{p^{2}(\overline{p}^{2}+\alpha
^{2}p^{4}/(4\Lambda ^{2}))}+\frac{\alpha }{2}\frac{\varepsilon _{\bar{\mu}%
\bar{\nu}\bar{\rho}}p_{\bar{\rho}}}{\overline{p}^{2}+\alpha
^{2}p^{4}/(4\Lambda ^{2})}.  \label{gapro}
\end{equation}
where $p^{4}$ stands for $(p^{2})^{2}$. It is not easy to use this
propagator for explicit computations, because of the structure of its
denominators. Away from the Faynman gauge the propagator is even more
involved.

\end{document}